\documentclass{article}

\usepackage{amsmath, amsfonts, amssymb}
\usepackage{xcolor}
\usepackage{graphicx}
\usepackage{hyperref}
\usepackage[a4paper,
    left=3.5cm,
    right=3.5cm,
    top=2.5cm,
    bottom=2.5cm
]{geometry}
\newcommand{\il}{\mathfrak{s}}
\newcommand{\iqq}{\mathfrak{b}}
\newcommand{\ipp}{\mathfrak{q}}
\newcommand{\iqqq}{q}
\newcommand{\qele}{e}
\newcommand{\ket}[1]{\left| #1 \right\rangle}
\newcommand{\bra}[1]{\left\langle #1 \right|}

\title{A Simple Understanding of Quantum Electrodynamics Using Bohmian Trajectories: Detecting Non-Ontic Photons}

\author{
Juan José Seoane$^{1}$, 
Abdelilah Benali$^{1}$ and
Xavier Oriols$^{1}$\thanks{Correspondence: \href{mailto:xavier.oriols@uab.es}{xavier.oriols@uab.es}} \\\small{$^1$Departament d’Enginyeria Electrònica, Universitat Autònoma de Barcelona (UAB)},\\ \small{Bellaterra, 08193 Barcelona, Spain}
}

\begin{document}
\maketitle

% =========================
% Abstract
% =========================
\begin{abstract}
The use of Bohmian mechanics as a practical tool for modeling non-relativistic quantum phenomena of matter provides clear evidence of its success, not only as a way to interpret the foundations of quantum mechanics, but also as a computational framework. In the literature, it is frequently argued that such a realistic view—based on deterministic trajectories—cannot account for phenomena involving the “creation” and “annihilation” of photons.
In this paper, by revisiting and rehabilitating earlier proposals, we show how quantum optics can be modeled using Bohmian trajectories for electrons in physical space, together with well-defined electromagnetic fields evolving in time.
By paying special attention to an experimental scenario demonstrating partition noise for photons, and to how the Born rule emerges in this context, the paper pursues two main goals. First, it validates the use of this simple Bohmian framework for pedagogical and computational purposes in understanding and visualizing quantum electrodynamics phenomena. Second, given that measurements are ultimately indicated on matter pointers, it clarifies what it means to measure photon or electromagnetic-field properties, even when they are considered non-ontic elements.
\end{abstract}

% =========================
% Keywords
% =========================

\section{Introduction}
\label{intro}

Quantum mechanics occupies a rather peculiar position in the history of science. 
On the one hand, many researchers focus on applying the formal rules of quantum theory to obtain quantitative predictions, 
while others are concerned with understanding what, if~anything, the~theory says about the underlying nature of physical reality. 
Several interpretations of quantum theory address this foundational question, 
among them are Bohmian mechanics~\cite{Bohm,Bell,Holland,Durr,BohmHiley,Kaloyerou1994,Travis_book}, the~many-worlds interpretation~\cite{Everett1957}, GRW~\cite{GRW1986} and~others.

In this context, Bohmian mechanics plays a distinctive role. 
Unlike most interpretations, which are mainly only invoked in foundational contexts, Bohmian mechanics has been widely applied to the modeling of non-relativistic phenomena involving quantum matter. Its ability to reproduce standard quantum predictions in this regime while offering an intuitive trajectory-based description has made it particularly attractive for both numerical simulations and conceptual analyses~\cite{book2019,Oriols2007,Oriols2016,book2023,Wyatt2005,SanzMiret2012,SanzMiret2014, SanzMiret2012_2, Deckert_semiclassical}.

Nevertheless, it is often argued that such a realist Bohmian view, grounded in deterministic particle trajectories, is fundamentally incompatible with processes involving the creation and annihilation of particles.
This objection is frequently regarded as one of the main conceptual limitations of the traditional formulation of Bohmian mechanics~\cite{Bell1986}. Several extensions of the Bohmian theory have been proposed in the literature to overcome this supposed limitation while preserving the core idea of the theory: namely, that there is always something in the world possessing well-defined properties (the so-called beable~\cite{Bell1986}), even in the absence of measurement. Quite different Bohmian-like theories are developed by considering different elements as~beables.

One possibility is to treat both fermions and bosons as beables in the form of point particles, whose trajectories may undergo stochastic creation and annihilation events. This idea was first suggested by Bell~\cite{Bell1986} and later developed in detail in the Bohmian quantum field theory models of D\"urr and collaborators~\cite{Durr2004,Durr2005}. An~alternative approach assigns beable status not to point particles but to fields. In~this case, the~fundamental beables are field configurations evolving deterministically in time. Variants of this idea can already be found in earlier work by Bohm for the electromagnetic field~\cite{Bohm1952II}. For~its more general application to either bosonic or fermionic fields, see the review in~\cite{Struyve}. 

Another possibility is to use a mixture of the previous beable assignments: employing beable-fields for bosons while preserving the original Bohmian beable-particles for fermions. The~mathematical machinery for treating fields as beables can be greatly simplified by using a mode decomposition, where the degrees of freedom needed to define the fields reduce to the coefficients representing the weight of each mode. In~this paper, by~revisiting older Bohmian proposals of this last type~\cite{Bohm1987,Belinfante1973,Holland,Kaloyerou1994}, we show that cavity quantum electrodynamics can be fully understood using traditional Bohmian trajectories for fermions, together with well-defined mode-decomposition coefficients for electromagnetic fields. The~practical solution just requires the traditional Bohmian trajectories guided by wave-function solutions of the unitary Schrödinger~equation.

The apparent conflict between employing well-defined trajectories, the solution of the unitary Schrödinger equation at all times for electrons and electromagnetic fields, and~describing photon creation and annihilation is largely a semantic problem rooted in the historical interpretation of photons. Once a photon is categorized as a \textit{particle}, it is implicitly granted some type of ontic status\footnote{Typically
, a~quantum system with $\mathcal{N}$ particles is described within a Hilbert space with $\mathcal{N}$ degrees of freedom. More generally, a~quantum system with a variable particle-number is formulated in Fock space~\cite{Sebens2022}}. Consequently, when a photon is absorbed, one is led to say that the photon, understood as a particle, has disappeared. However, the~process of photon creation and annihilation can be reinterpreted more naturally by avoiding the term \textit{particle} and defining the photon instead as a quantized amount of energy of the electromagnetic field. From~this perspective, increases or decreases in the field’s energy are straightforwardly described within a wave-function solution of the ordinary Schrödinger~dynamics.

Using this mixed ontology, the~present work studies scenarios where the measurement apparatuses are included as part of the dynamical system. By~introducing pointer degrees of freedom, the~effective configuration space of the system is enlarged. This allows the different branches of the wave function to become dynamically separated, leading to an effective collapse of the wave function in the Bohmian sense. Within~this framework, we show, through concrete numerical examples, how phenomena typically interpreted in terms of particle-like photons can be understood entirely in terms of fermionic trajectories interacting with quantized electromagnetic fields. In~particular, we analyze photon partition noise between two spatially separated detectors, where an incoming photon is registered by one detector or the other, but~not both—a phenomenon often invoked as evidence of wave–particle (‘Jekyll-and-Hyde’ \cite{Catani2023}) photon duality. However, we show that the observed discreteness in the experimental results can be explained by the particle-like nature of the material detectors rather than by a particle ontology of the electromagnetic field\footnote{It is relevant to mention that Goldstein~et~al.~\cite{Goldstein2007} explored the possibility that not all particles need to be real; in their approach, certain degrees of freedom, such as those associated with radiation, are not assigned positions as beables, yet the theory still reproduces all empirical predictions. The~numerical results in this paper can be understood as a concrete example of the result of~\cite{Goldstein2007}.}.

From these results, where the measuring pointers are made of matter, it seems that the measurement of photons, or~even electromagnetic fields, in~Bohmian mechanics can be explained by considering only matter as beable elements. Therefore, the~possibility of an ontology for quantum electrodynamics, different from the previously mentioned ones, appears viable. The~present paper provides arguments in favor of this ontology in which fermions are beable point particles described by trajectories, while electromagnetic fields are emergent properties arising from other non-simulated ontic fermions that generate such fields. This idea has been already anticipated by some authors. {For example, an~approach where the ``Dirac sea is taken seriously'', in~terms of assigning ontological status only to fermions, was proposed by Colin and Struyve~\cite{ColinStruyve} in which all fermionic degrees of freedom have well-defined trajectories. See similar proposals in Refs.
.~\cite{Dustin2018, Deckert_thesis}. A significant inspiration for all these  “fermions-only” ontology comes from earlier attempts to formulate classical electrodynamics without fields, relying solely on retarded action-at-a-distance interactions between point-particles
~\cite{Schwarzschild1903,Tetrode1922,Fokker1929,Feynman1945,Feynman1949}.

An important clarification is that the present work is restricted to a non‑relativistic description. Although~Maxwell's equations are fundamentally relativistic, cavity quantum electrodynamic experiments are routinely and successfully described using non‑relativistic kinetic energy for matter.  A~fully relativistic formulation of the present model would require substantial additional theoretical development and lies beyond the scope of this work\footnote{A possible direction for extending this proposal to relativistic scenarios is indicated in Lienert~\cite{Lienert2018} where particles interact directly along light cones at the quantum level, using multi-time wave functions to maintain relativistic consistency.}.

The structure of the paper is as follows. After~this introduction, we discuss the historical meaning of electromagnetic fields in Section~\ref{intro1} and photons in Section~\ref{intro2}. Then, we discuss how photon and light properties can be experimentally measured within Bohmian mechanics in Section~\ref{intro3}. In~Section~\ref{sec2} we develop the classical light–matter Hamiltonian, mentioning the mode decomposition in Section~\ref{sec24}. Next, in~Section~\ref{sec3}, the~canonical quantization of both the matter degree of freedom in Section~\ref{sec31} and field degrees of freedom in Section~\ref{sec32} are discussed, leading to the Schrödinger equation for the full light–matter system in Section~\ref{sec33}. A~Bohmian reformulation of the theory is subsequently introduced in Section~\ref{sec4}. In~Section~\ref{sec5} we present a discussion of the Born rule and how the standard statistical predictions are recovered. Numerical simulations of two electrons interacting with quantized light, with~and without the measuring apparatuses, are presented in Section~\ref{sec6} to discuss the detection of non-ontic photons, and~the paper concludes with a summary in Section~\ref{conc}. Several appendices discuss technical~details.

%%%%%%%%%%%%%%%%%%%%%%%%%%%%%%%%%%%%%%%%%%%%%%%%%%%%%%%%%%%%%

\subsection{What Are Electromagnetic Fields?}
\label{intro1}

In elementary physics courses, when dealing with two charged particles, we can define the electric field as the force generated by the first particle at the location of the second particle when the latter is replaced by a unit positive test charge. The~usefulness of the concept of the electric field is that we do not need to explicitly acknowledge the presence of the first particle to describe the dynamics of the second particle. Considering only the second particle and the electric field is sufficient to describe its~dynamics. 

This definition allows us to regard the electromagnetic field as a convenient effective (emergent) tool for describing forces among charged particles when not all particles are explicitly included in the discussion. In~this view, the~field is not something ontic (that physically exists in reality), but~an emergent one\footnote{Certainly, the~word ontic can have slightly different meanings. Even a law can be defined as a nomological element that is part of the ontology of a theory. In~this paper, however, we will use the term "ontic element of a physical theory" to refer to something that is considered physically real, in~the same sense that particles are considered in classical mechanics. For~an element of a physical theory that is not considered physically real in this sense, we will refer to it as non-ontic.  For~example, in~classical mechanics, the~center of mass—defined as a sum of the positions of the particles—is not ontic, even though the particles themselves are ontic. In~the literature, what we call ontic in this paper is often referred to as the primitive ontology of the theory~\cite{Allori2008}.}. However, the~electromagnetic field can also be understood as a physically real entity that exists in space independently of charged particles. In~this latter view, one assumes that the electromagnetic field is something physical, that is, something~ontic.

The earliest descriptions of electrodynamics were formulated according to the first non-ontic viewpoint: the electromagnetic field was something emergent from the particles. Coulomb~\cite{Coulomb} described electrostatic interactions through direct forces among charges, and Amp\`ere extended this view to moving charges and electric currents~\cite{Ampere}. Weber later proposed a unified action-at-a-distance theory of electricity and magnetism~\cite{Weber}. It is interesting to emphasize that, in~this view, the~electric field ``seen'' by one particle is different from the electric field "seen" by another particle. Therefore, there is no single, universal electric field, but~rather a set of effective electric fields, each one associated with each particle, which determines how a given particle  moves under the influence of all the~others.

Faraday was perhaps the first to attribute a kind of physical reality (rather than merely mathematical significance) to electromagnetic fields when he introduced the idea of continuous “lines of force” \cite{Faraday}. Maxwell later formalized this vision, defining fields as local physical quantities whose variations propagate as electromagnetic waves~\cite{Maxwell}. Notice that Maxwell’s equations are written for a single electromagnetic field. Thus, one can speak of the ontic electromagnetic field. Since Maxwell’s equations predicted the propagation of electromagnetic disturbances at the finite speed of light, the~instantaneous action-at-a-distance picture of Weber and others was gradually abandoned, and~the ontic conception of the electromagnetic field became the standard framework of classical~electrodynamics.

The first quantum formulation of electrodynamics, developed by Dirac, was based on Maxwell’s ontological view of a single electromagnetic field~\cite{DiracQED} where the~degrees of freedom that define the electromagnetic fields (written in terms of canonical variables) were quantized. As~a result, electromagnetic energy---like the energy of quantum matter---was found to be~quantized.

Interestingly, formulations of electrodynamics without ontic electromagnetic fields were never completely abandoned. In a classical framework, Schwarzschild~\cite{Schwarzschild1903}, Tetrode~\cite{Tetrode1922}, and~Fokker~\cite{Fokker1929} addressed the problems of Weber’s instantaneous action-at-a-distance picture and developed relativistically consistent retarded action-at-a-distance theories without postulating an ontic electromagnetic field. The~main practical advantage of these approaches is the absence of the problems related to self-interaction and the associated infinities~\cite{Dustin2018}. 

Later, Wheeler and Feynman proposed a formulation of classical electrodynamics in which charged particles interact directly through advanced and retarded interactions~\cite{Feynman1945,Feynman1949}. Feynman’s later path-integral formulation of quantum electrodynamics was partly inspired by this line of thought~\cite{FeynmanQED}.  {{More recently, Lienert~\cite{Lienert2018} proposed a quantum version of this idea, in~which particles interact directly along light cones at the quantum level, using multi-time wave functions to maintain relativistic consistency.}}  %%%%%%%%%%%%%%%%%%%%%%%%%%%%%%%%%%%%%%%%%%%%%%%%%%%%%%%%%%%%%

\subsection{What Are Photons?}
\label{intro2}

We have seen that, independently of whether light is assigned an ontic or merely a mathematical status,  the~electromagnetic field(s) alters the dynamics of particles by transferring energy and momentum among them in discrete (quantized) amounts. 

The idea that the energy of light comes in discrete packets was introduced by Planck in 1900 in his study of blackbody radiation~\cite{Planck}. 
In 1905, Einstein proposed that light itself consists of individual energy quanta in order to explain the photoelectric effect~\cite{Einstein}. It was therefore natural to interpret the electromagnetic field as being composed of a type of light particle that exchanges energy and momentum with matter particles. These light particles were later named photons by Lewis~\cite{Lewis}. These early models treated photons as discrete ontic particles with fixed energy and momentum and were able to explain, for~example, the~photoelectric effect\footnote{One can imagine photons and electrons as billiard balls satisfying conservation of energy and momentum when colliding. In~the process, the~photon can be created or annihilated.) However, this simple model cannot account for interference and other intrinsically quantum electrodynamic phenomena~\cite{Catani2023}.}.

Historically, a~complete and accurate description of photons emerged through the quantum field theory, specifically the quantum electrodynamics~\cite{QED,Banks2008,Loudon,Scully,Peskin}. In~this framework, the~electromagnetic field undergoes a quantization procedure, analogous to the one applied to electrons. A~quantum electron can be understood as a superposition of classical positions, each with a certain probability. Similarly, a~quantum electromagnetic field can be viewed as a  superposition of classical electromagnetic fields, each weighted by a probability amplitude.
In the traditional language of the quantum field theory, the~photon is neither a wave nor a particle with well-defined positions, but~an excitation of the quantum electromagnetic field\footnote{The word \textit{particle} is commonly used in quantum field theory. It certainly does not mean an entity with well-defined positions at all time, but~something different and compatible with its definition of \textit{excitation of the quantum electromagnetic field}. In~any case, the~word \textit{particle} for the photon still suggests some type of onticity for the photon in the sense of an additional dimension in Hilbert (or Fock) space.}. Mathematically, such an excitation is created when a creation operator acts on a quantum state of the field. We then say that a photon has been created.
Because photon creation and annihilation can occur in various individual scenarios, and the principle of quantum superposition still holds, a quantum field state is generally a superposition of states with different numbers of photons. This means that the number of particles is not fixed.  Consequently, the~standard Schrödinger equation, which ensures conservation of probability for a  fixed number of particles, cannot be directly applied.
To handle processes involving  variable particle numbers , a~new structure called  Fock space  is required. Fock space allows one to describe quantum states with  different numbers of particles simultaneously  (see Appendix~\ref{appendix}).

However, this quantum-field-theoretic machinery of creating and annihilating photons can also be understood as a “way of talking” if the photon is interpreted not as a particle created or destroyed, but~merely as a quantized amount of energy created or destroyed. In~that case, “creating a photon” can be translated as “increasing the energy of the electromagnetic field by a quantized amount.” There is then no need to invoke a new ontic entity to describe an increment in the energy of the electromagnetic field. This will be the view followed in this paper. We can still refer to this increment of energy as photon, but~there is nothing ontic in this new definition. In~this new “way of talking,” neither matter nor light appears or disappears; rather, they only exchange energy between them. We will see that the (unitary) Schrödinger equation remains a valid equation of motion to describe the joint evolution of light and matter and how they exchange energy in the quantum~regime.

%%%%%%%%%%%%%%%%%%%%%%%%%%%%%%%%%%%%%%%%%%%%%%%%%%%%%%%%%%%%%

\subsection{Measuring Photon and Light~Properties}
\label{intro3}

The assumption defended in this paper---that photons are not ontic particles---raises what might appear to be a paradoxical question: how can a measurement detect a photon at a given position if photons themselves are not ontic elements? The answer is that not everything that is measured must correspond to an ontic element. What is ontic is defined by the theory, not by what can be measured in an~experiment. 

For example, one can measure the temperature, but~in Bohmian and orthodox quantum mechanics there is no ontic element called temperature; it is an emergent property arising from the dynamics of many particles. Similarly, measuring the spin and getting a number does not imply that the spin is an ontic property.  For~example, before~the measurement, the~spin can be in a superposition of spin up and spin down without a well-defined up or down value. Another example is the Bohmian velocity, which can be measured in the laboratory through weak values~\cite{Kocsis2011,Oriols2025}. In~the orthodox theory, such an empirical number represents only a statistical value, without~any ontic status. In~Bohmian mechanics, however, despite the number being obtained after statistical post-selection, it corresponds directly to the ontic velocity of the~particles.

Consequently, there is not a conceptual problem in defending that measured properties usually attributed to photons—such as energy, momentum, or~polarization—do not correspond to intrinsic attributes of ontic photons.  A~photo detector click corresponds to a change in the configuration of material particles (typically electrons) whose motion is influenced by the electromagnetic fields. In~fact, one can also assume that this electromagnetic field is an emergent property created by some emitter electrons. Thus, if~we want, we can argue that what the photo detector is measuring is a correlation between emitter and detector matter ontic particles\footnote{A completely different approach, in~which the ontic variables are fields rather than particles and the detected entities are ontic fields, can also be defended~\cite{Struyve2006}.}.  

In fact, the~(orthodox) quantum field theory still has the problem of the measuring collapse inherited by ordinary quantum mechanics~\cite{Bell1990, Maudlin1995}. The~measurement collapse problem is the tension between the linear, deterministic, and~unitary evolution of the quantum (or quantum field) state and the postulated non-linear, stochastic “collapse” to a single outcome during measurement, which has no clear physical explanation within neither the (orthodox) quantum theory nor (orthodox) quantum field theory. This problem simply disappears in our Bohmian explanation, as~will be seen in Section~\ref{sec5}.

As discussed previously, we defend that photons can be interpreted as mere quantized amounts of energy rather than as genuine ontic particles. The~crucial point is that all measurements in the laboratory ultimately reduce to measurements of the positions of material particles. 
In particular, we show that the photon partition noise is not a manifestation of its particle nature, but~it can be understood perfectly well as just an artifact of the detection process with matter pointers described by ontic particles. This ability of Bohmian mechanics to explain all types of measurements with matter pointers suggests a compelling strategy for maintaining only an ontology for~matter.

%%%%%%%%%%%%%%%%%%%%%%%%%%%%%%%%%%%%%%%%%%
\section{Classical Light--Matter~Hamiltonian}
\label{sec2}

First of all we clarify whether the meaning of degrees of freedom and the meaning of an ontic property are the same or not.  A~degree of freedom is a mathematical element used to describe a particular state of a system. By~contrast, an~ontic property is an element that the theory identifies as corresponding to something physically real. In~classical mechanics, for~example, the~position of a particle is an ontic property and, at~the same time, it can serve as a degree of freedom. However, in~certain problems, we may describe a system of two particles, with~two ontic positions, using the center of mass of the two positions as a new degree of freedom; in this case, the~center of mass becomes a useful mathematical degree of freedom, but~it is not itself a primitive ontic property. As~we have discussed in subsection \ref{intro1}, several attempts~\cite{Schwarzschild1903,Tetrode1922,Fokker1929,Dustin2018,Feynman1945,Feynman1949} have attributed to the classical electromagnetic field a similar role: it can be assumed as a degree of freedom encapsulating the role of non-simulated ontic particles that generate such electromagnetic~field. 

\subsection{Matter}
\label{sec21}

Let us start by considering a system of $\mathcal{N}$ matter particles (typically electrons) with masses $m_e$ and charges $\qele$. The~degrees of freedom of the system are the particle positions
 $\mathbf{r}_i$, with~$i=1,\dots,\mathcal{N}$ in~three-dimensional physical space and their corresponding (canonical) momenta being $\mathbf{p}_i=m_e \dot{\mathbf{r}}_i(t)$ with $\dot{\mathbf{r}}_i(t)$ as the velocity of the electron. Together, these variables define the $6\mathcal{N}$-dimensional phase space of the system, which will be used to construct the Hamiltonian\footnote{As time evolves, in~classical mechanics, the~system follows a trajectory in the $6\mathcal{N}$-dimensional phase space that corresponds to a set of trajectories $\mathbf{r}_i(t)$ in physical space, one for each particle. These trajectories are regarded as ontic elements of the classical mechanics theory, describing the actual motion of matter.}.

The charge density $\rho(\mathbf{r},t)$ and current density $\mathbf{J}(\mathbf{r},t)$ describe the distribution and motion of electric charge at the position $\mathbf{r}$ of the physical space\footnote{Why are we interested in knowing what happens at $\mathbf{r}$ if, for~the moment, no particles are present there? We provide an answer when we discuss the retarded potentials in \eqref{retardedpotentialsca} and \eqref{retardedpotentialvec}.}. At~time $t$, due to the presence of a collection of charged particles at positions $\mathbf{r}_i(t)$:
\begin{equation}
\rho(\mathbf{r},t) = \sum_{i=1}^{\mathcal{N}} \qele \, \delta(\mathbf{r} - \mathbf{r}_i(t)),
\label{chargeden}
\end{equation}
\begin{equation}
\mathbf{J}(\mathbf{r},t) = \sum_{i=1}^{\mathcal{N}} \qele \, \dot{\mathbf{r}}_i(t) \,
\delta(\mathbf{r} - \mathbf{r}_i(t)).
\label{currentden}
\end{equation}
The charge \eqref{chargeden} and current  densities \eqref{currentden} satisfy the  continuity equation
\begin{equation}
\frac{\partial \rho(\mathbf{r},t)}{\partial t}
+ \nabla \cdot \mathbf{J}(\mathbf{r},t) = 0,
\label{continuitysingle}
\end{equation}
which expresses the local charge conservation of~matter.

\subsection{Electromagnetic Field(s)}
\label{sec22}

We will indistinctly use the name ``light'' or the name ``electromagnetic fields'' to define two vector fields, the~electric $\mathbf{E}(\mathbf{r},t)$ and the magnetic $\mathbf{B}(\mathbf{r},t)$ fields,  in~three-dimensional physical space $\mathbf{r}$ and time $t$.  These (gauge-invariant) electromagnetic fields can be expressed in terms of the (gauge-dependent) scalar potential $A_0(\mathbf{r},t)$ and vector potential $\mathbf{A}(\mathbf{r},t)$ as
\begin{equation}
\mathbf{E}(\mathbf{r},t) = - \nabla A_0(\mathbf{r},t) 
- \frac{\partial \mathbf{A}(\mathbf{r},t)}{\partial t},
\label{defE}
\end{equation}
and
\begin{equation}
\mathbf{B}(\mathbf{r},t) = \nabla \times \mathbf{A}(\mathbf{r},t).
\label{defB}
\end{equation}
From these definitions, two of Maxwell's equations are automatically satisfied. First, since the divergence of a curl vanishes, we get $\nabla \cdot \mathbf{B}(\mathbf{r},t) 
= \nabla \cdot (\nabla \times \mathbf{A}) = 0$ which expresses the absence of magnetic monopoles. Second,
\begin{equation}
\nabla \times \mathbf{E}(\mathbf{r},t)
= - \nabla \times \nabla A_0 
- \frac{\partial}{\partial t} (\nabla \times \mathbf{A})
= - \frac{\partial \mathbf{B}(\mathbf{r},t)}{\partial t},
\label{nablae}
\end{equation}
 which is Faraday's law of induction. The~remaining two Maxwell equations relate the electromagnetic fields to the charge \eqref{chargeden} and current densities \eqref{currentden}, as
\begin{equation}
\nabla \cdot \mathbf{E}(\mathbf{r},t) 
= \frac{\rho(\mathbf{r},t)}{\varepsilon_0},
\label{gausslaw}
\end{equation}
which is known as Gauss's law, and~\begin{equation}
\nabla \times \mathbf{B}(\mathbf{r},t) 
= \mu_0 \mathbf{J}(\mathbf{r},t) 
+ \mu_0 \varepsilon_0 
\frac{\partial \mathbf{E}(\mathbf{r},t)}{\partial t},
\label{ampere}
\end{equation}
which is known as the Ampère (or Ampère–Maxwell) law. At~this point it seems that the electromagnetic fields are something real (ontic), at~least, as~real as the particles. For~sure, as~discussed in Section \ref{intro1} this is one possibility widely extended in the treatment of classical electrodynamics. However, as~discussed also in the Sections \ref{intro1} and \ref{intro2}, this is not the only~possibility.   

In the Lorenz gauge, the~electromagnetic potentials can be shown to be solution of an inhomogeneous wave equation that only depends on the charge and current densities defined in \eqref{chargeden} and \eqref{currentden}. {Under the assumption that no electromagnetic radiation is incoming from infinity, the~solutions of these equations are the retarded potentials} \cite{Jackson}:
\begin{align}
A_0(\mathbf{r},t) 
&= \frac{1}{4\pi \varepsilon_0} 
\int \frac{\rho(\mathbf{r}', t_r)}
{|\mathbf{r}-\mathbf{r}'|} \, d^3 r'  = \sum_{i=1}^{\mathcal{N}}\frac{1}{4\pi \varepsilon_0} 
\int \frac{ \qele \, \delta(\mathbf{r}' - \mathbf{r}_i(t_r))}
{|\mathbf{r}-\mathbf{r}'|} \, d^3 r' \nonumber\\
&= \sum_{i=1}^{\mathcal{N}}\frac{\qele}{4\pi \varepsilon_0} 
\frac{1}{|\mathbf{r}-\mathbf{r}_i(t_{r,i})|}.
\label{retardedpotentialsca}
\end{align}
and
\begin{align}
\mathbf{A}(\mathbf{r},t) 
&= \frac{\mu_0}{4\pi}
\int \frac{\mathbf{J}(\mathbf{r}', t_r)}
{|\mathbf{r}-\mathbf{r}'|}
\, d^3 r'  = \sum_{i=1}^{\mathcal{N}} \frac{\mu_0}{4\pi}
\int \frac{ \qele \, \dot{\mathbf{r}}_i(t_r)
\, \delta(\mathbf{r}' - \mathbf{r}_i(t_r))}
{|\mathbf{r}-\mathbf{r}'|}
\, d^3 r' \nonumber\\
&= \sum_{i=1}^{\mathcal{N}} \frac{\mu_0}{4\pi} \qele \,
\frac{\dot{\mathbf{r}}_i(t_{r,i})}
{|\mathbf{r}-\mathbf{r}_i(t_{r,i})|}.
\label{retardedpotentialvec}
\end{align}
where the retarded time\footnote{Similarly, an~advanced time can also be mathematically defined for the solution of the mentioned inhomogeneous equation. For~a discussion of its physical meaning in classical scenarios, see~\cite{Feynman1945,Feynman1949}.} is defined as $t_r = t - \frac{|\mathbf{r}-\mathbf{r}'|}{c}$ and $t_{r,i} = t - \frac{|\mathbf{r}-\mathbf{r}_i(t_r)|}{c}$. These expressions for the retarded potentials show explicitly that the electromagnetic potentials (and therefore the fields) at $\mathbf{r}$ are completely determined by the charge and current densities defined in \eqref{chargeden} and \eqref{currentden}, from~the set of particles $i=1,\ldots,\mathcal{N}$, as~anticipated a long time ago by Schwarzschild~\cite{Schwarzschild1903}, Tetrode~\cite{Tetrode1922}, and~Fokker~\cite{Fokker1929}.

Therefore, for~an additional particle identified with the label $\mathcal{N}+1$ and located at $\mathbf{r}=\mathbf{r}_{\mathcal{N}+1}$, the~electromagnetic fields $\mathbf{E}(\mathbf{r}_{\mathcal{N}+1},t)$ and $\mathbf{B}(\mathbf{r}_{\mathcal{N}+1},t)$ can thus be regarded as emergent quantities, arising from the positions and velocities of the set of other particles $i=1,\ldots,\mathcal{N}$, through \eqref{retardedpotentialsca} and \eqref{retardedpotentialvec}.  The~final equation of motion of the $\mathbf{r}=\mathbf{r}_{\mathcal{N}+1}$ electron is the well-known Lorentz force:
\begin{equation}
m_{e}\,\ddot{\mathbf{r}}_{\mathcal{N}+1}(t)
=
q_{e}
\left[
\mathbf{E}(\mathbf{r}_{\mathcal{N}+1}(t),t)
+
\dot{\mathbf{r}}_{\mathcal{N}+1}(t)
\times
\mathbf{B}(\mathbf{r}_{\mathcal{N}+1}(t),t)
\right].
\label{lorentzforceNplus1}
\end{equation}
Notice that the dynamics of the particle labeled as $\mathbf{r}=\mathbf{r}_{\mathcal{N}+1}$ depends only on the ontic positions of the other $\mathcal{N}$ particles through \eqref{retardedpotentialsca} and \eqref{retardedpotentialvec}. 

There is nothing special about the particle labeled $\mathcal{N}+1$. By~the same reasoning, we would need another set of electromagnetic retarded potentials to compute the dynamics of particle $i=1$. Such potentials are obtained from \eqref{retardedpotentialsca} and \eqref{retardedpotentialvec} but using the charge and current densities defined in \eqref{chargeden} and \eqref{currentden}, summing over the particles $i=2, 3, \ldots, \mathcal{N}+1$, which amounts to excluding the interaction of particle $1$ with itself, corresponding to self-interaction. The~same reasoning applies to particles $2$, $3$, and~so on. Thus, one needs as many electromagnetic fields (or potentials) as particles~\cite{Schwarzschild1903,Tetrode1922,Fokker1929,Dustin2018,Feynman1945,Feynman1949}. This is the price to pay if one wants to treat matter as an ontic element while regarding the classical electromagnetic fields or potentials as convenient emergent~entities.

In the rest of this section (Section~\ref{sec2}), we develop only the Hamiltonian for the electron at $\mathbf{r} = \mathbf{r}_{\mathcal{N}+1}$ interacting with the electromagnetic fields discussed above. This one-electron Hamiltonian admits two possible~interpretations.

The first interpretation is that the whole system contains ${\mathcal{N}+1}$  ontic electrons, where the interaction of the $\mathcal{N}$ other ontic electrons with the electron labeled by $\mathbf{r} = \mathbf{r}_{\mathcal{N}+1}$ is encapsulated in the electromagnetic fields, which emerge as non-ontic degrees of freedom and are computed from Equations~\eqref{retardedpotentialsca} and \eqref{retardedpotentialvec}. A~complete treatment within this interpretation would therefore require $\mathcal{N}$ additional Hamiltonians, one for each of the remaining electrons. This first interpretation is not further developed in this~paper.

A much simpler second interpretation is to adopt the mixed ontology introduced in the Introduction, where the system consists of a single ontic electron and an ontic electromagnetic field. This approach allows us to reuse the standard quantum optics machinery to canonically quantize the classical Hamiltonian in Section~\ref{sec3}. The~numerical results presented in Section~\ref{sec6} are based on this mixed ontology.  Therefore, strictly speaking, the~results in this paper, Section~\ref{sec6}, only demonstrate that quantum cavity electrodynamics can be fully understood without ontic photons, but~they do not establish that the same phenomena can be explained without ontic electromagnetic~fields.

\subsection{Minimal Coupling~Hamiltonian}
\label{sec23}

The total Hamiltonian for one electron interacting with the electromagnetic field (ontic or created by the rest of the electrons not explicitly simulated) reads as
\begin{equation}
H = \frac{1}{2m_e}
\left(
\mathbf{p} - \qele \mathbf{A}(\mathbf{r},t)
\right)^2 + \qele A_0(\mathbf{r},t)+H_{\text{field}},
\label{H_total}
\end{equation}
where the electromagnetic field energy is
\begin{equation}
H_{\text{field}} =
\frac{\varepsilon_0}{2}
\int d^3 r \,
\left[
\mathbf{E}_\perp^2(\mathbf{r},t)
+ c^2 \mathbf{B}^2(\mathbf{r},t)
\right].
\label{energy1}
\end{equation}
It is interesting to discuss here the argument presented in most textbooks on quantum optics~\cite{Grynberg2011,Cohen,Scully} on why only the energy of the {transversal} component is considered. The~electric field is split into longitudinal and transverse components: $\mathbf{E}(\mathbf{r},t) = \mathbf{E}_L(\mathbf{r},t) + \mathbf{E}_\perp(\mathbf{r},t)$. Gauss's law \eqref{gausslaw} only involves the longitudinal component because $\nabla \cdot \mathbf{E}_\perp(\mathbf{r},t)=0$ by construction. Thus, since the charge density in \eqref{chargeden} is determined by the particle positions $\mathbf{r}_i$, the~longitudinal electric field is not an independent dynamical degree of freedom. In~the Coulomb gauge, the~electromagnetic (Coulomb) energy of the longitudinal field is encapsulated in the scalar potential $A_0(\mathbf{r},t)$, which depends on all matter positions. This is exactly the argument we have invoked when writing \eqref{retardedpotentialsca}. Although we do not develop this further in this paper, we suggest that the same argument applied to $A_0(\mathbf{r},t)$ can also be applied to $\mathbf{A}(\mathbf{r},t)$ through \eqref{retardedpotentialvec}.

\subsection{Mode~Decomposition}
\label{sec24}

By using \eqref{nablae} and \eqref{ampere}, the~transverse part contains the dynamical, propagating degrees of freedom of the field, satisfying
\begin{equation}
\nabla^2 \mathbf{E}_\perp(\mathbf{r},t) - \frac{1}{c^2}\frac{\partial^2 \mathbf{E}_\perp(\mathbf{r},t)}{\partial t^2} 
= \mu_0 \frac{\partial \mathbf{J}_\perp(\mathbf{r},t)}{\partial t}.
\label{mode1}
\end{equation}
and putting $\mathbf{E}_\perp(\mathbf{r},t)= -\frac{\partial \mathbf{A}(\mathbf{r},t)}{\partial t}$ into \eqref{mode1},  we finally get
\begin{equation}
\nabla^2 \mathbf{A}(\mathbf{r},t)
- \frac{1}{c^2}\frac{\partial^2 \mathbf{A}(\mathbf{r},t)}{\partial t^2}
= -\,\mu_0 \mathbf{J}_\perp(\mathbf{r},t).
\label{mode3}
\end{equation}
To perform a mode decomposition, we expand the transverse vector potential in terms of a complete set of orthonormal  functions $\mathbf{U}_\lambda(\mathbf{r}) \in \mathbb{R}^3$ that depend only on the position,  $\int_{\infty}^{\infty} d\mathbf{r} \; \mathbf{U}_\lambda'(\mathbf{r})\mathbf{U}_\lambda(\mathbf{r})=\delta_{\lambda,\lambda'}$. Then, by~defining a time-dependent amplitude as \linebreak  $\il_\lambda(t)=\int_{\infty}^{\infty} d\mathbf{r} \; \mathbf{A}(\mathbf{r},t)\mathbf{U}_\lambda(\mathbf{r})$ we get
\begin{equation}
\mathbf{A}(\mathbf{r},t) = \sum_{\lambda=1}^\mathcal{M} \il_\lambda(t) \, \mathbf{U}_\lambda(\mathbf{r}),
\label{expansion}
\end{equation}
We consider only a finite number $\mathcal{M}$ of modes\footnote{The units of $\il_\lambda(t)$ are the square root of a volume multiplied by volt-second and divided by meter.}. This mode decomposition in \eqref{expansion} divides \eqref{mode3} into a temporal and a spatial equations. For~practical reasons that will be evident later, it is convenient to use the following Helmholtz equation\footnote{If refractive index $n(\mathbf{r})$ varies spatially, the~equation remains $\nabla^2 \mathbf{U}_\lambda(\mathbf{r}) + \frac{\omega_\lambda^2}{c^2} n^2(\mathbf{r})
\mathbf{U}_\lambda(\mathbf{r})= 0,$ which is an eigenvalue problem for $\omega_\lambda$.} as the spatial equation:
\begin{equation}
\nabla^2 \mathbf{U}_\lambda(\mathbf{r}) + k_\lambda^2 \mathbf{U}_\lambda(\mathbf{r}) = 0.
\label{helmotz}
\end{equation}
In the Coulomb gauge, we also have $\nabla \cdot \mathbf{U}_\lambda(\mathbf{r}) = 0$. 
Substituting the expansion \eqref{expansion} into \eqref{mode3} and using the orthonormality of the mode functions yields a set of driven harmonic oscillator equations for the mode amplitudes:
\begin{equation}
\ddot{\il}_\lambda(t) + \omega_\lambda^2 \il_\lambda(t) = f_\lambda(t),
\end{equation}
with  $f_\lambda(t) = c^2 \mu_0 \int d^3 r \, \mathbf{U}_\lambda(\mathbf{r}) \cdot  \mathbf{J}_\perp(\mathbf{r},t)$, where we have defined $\omega_\lambda = c k_\lambda$. From~\eqref{expansion}, the~transverse electric field can be written as
\begin{equation}
\mathbf{E}_\perp(\mathbf{r},t) = - \frac{\partial \mathbf{A}(\mathbf{r},t)}{\partial t} 
= - \sum_{\lambda=1}^\mathcal{M} \dot{\il}_\lambda(t) \, \mathbf{U}_\lambda(\mathbf{r}).
\end{equation}
Equivalently, the~magnetic field can then be written in terms of the mode functions as
\begin{equation}
\mathbf{B}(\mathbf{r},t) = \nabla \times \mathbf{A}(\mathbf{r},t) 
= \sum_{\lambda=1}^\mathcal{M} \il_\lambda(t) \, \nabla \times \mathbf{U}_\lambda(\mathbf{r}).
\end{equation}

We have been able to describe the electromagnetic fields as a function of a set of parameters $\il_\lambda(t)$ plus functions $\mathbf{U}_\lambda(\mathbf{r})$. Indeed, the~ parameters $\il_\lambda(t)$ are the new degrees of freedom that describe the electromagnetic fields.  In~other words, $\il_\lambda(t)$ can take any value depending on the scenario, while the functions $\mathbf{U}_\lambda(\mathbf{r})$ are, by~construction, fixed by the geometry in \eqref{helmotz} and cannot vary in a given experiment\footnote{At this point the reader can argue that such new degrees of freedom, $\il_1(t),\il_2(t), \ldots$, used to describe the electromagnetic fields are somehow arbitrary. By~selecting different  $\mathbf{U}_\lambda(\mathbf{r})$ we will have different degrees of freedom. This is true and it would be problematic if we pretend to give some type of ontological status to the photons, but~this is not the case in this paper. The~analogy with the center of mass for describing particles is pertinent here. The~degree of freedom of the center of mass defined is not an ontic element, but~it can be used as a useful degree of freedom to solve the problem.}. Let's write the Hamiltonian with this new degree of~freedom. 

\subsubsection{Electromagnetic~Energy}

Finally, the~energy of the electromagnetic field in \eqref{energy1} written as the sum of the electrical and magnetic field energies, $H_{\text{field}}$, is given by
\begin{equation}
H_{\text{field}} =
\frac{\varepsilon_0}{2}
\int d^3 r \,
\left[
\mathbf{E}_\perp^2(\mathbf{r},t)
+ c^2 \mathbf{B}^2(\mathbf{r},t)
\right]=\frac{\varepsilon_0}{2} \sum_{\lambda=1}^\mathcal{M} \left( |\dot{\il}_\lambda(t)|^2+ \omega_\lambda^2 |\il_\lambda(t)|^2\right).
\label{energy2}
\end{equation}
where we used the orthonormality of the mode functions\footnote{Multiply the Helmholtz equation, \mbox{Equation \eqref{helmotz}}, for $\mathbf U_\lambda$ by
$\mathbf U_{\lambda'}^*$ and integrate $\int d^3 r \,\mathbf U_{\lambda'}^* \cdot\nabla^2 \mathbf U_\lambda+k_\lambda^2\int d^3 r \,\mathbf U_{\lambda'}^* \cdot\mathbf U_\lambda= 0$;
integrating by parts and using the boundary conditions, $\int d^3 r \,\mathbf U_{\lambda'}^* \cdot\nabla^2 \mathbf U_\lambda=\int d^3 r \,\nabla^2 \mathbf U_{\lambda'}^* \cdot\mathbf U_\lambda$; 
using the Helmholtz equation, Equation \eqref{helmotz}, again, we obtain $- k_{\lambda'}^2 \int d^3 r \,\mathbf U_{\lambda'}^* \cdot\mathbf U_\lambda+k_\lambda^2\int d^3 r \,\mathbf U_{\lambda'}^* \cdot\mathbf U_\lambda= 0$; therefore, $(k_\lambda^2 - k_{\lambda'}^2)\int d^3 r \,\mathbf U_{\lambda'}^* \cdot\mathbf U_\lambda= 0$; finally, if~$k_\lambda \neq k_{\lambda'}$, it follows that $
\int d^3 r \,\mathbf U_{\lambda'}^*(\mathbf r)
\cdot\mathbf U_\lambda(\mathbf r)= 0$, thus the modes are orthogonal; we properly normalize  $\int d^3 r \,\mathbf U_\lambda \cdot  \mathbf U^*_{\lambda} = 1$ so that  $ \int d^3 r \,\mathbf U_\lambda \cdot  \mathbf U^*_{\lambda'} =  \delta_{\lambda,\lambda'}$} and $\int d^3 r \, (\nabla \times \mathbf U_\lambda) \cdot (\nabla \times \mathbf U^*_{\lambda`}) =  k_\lambda^2 \delta_{\lambda,\lambda'}$ for normalized transverse modes\footnote{We compute $\int d^3 r \, (\nabla \times \mathbf U_\lambda) \cdot (\nabla \times \mathbf U^*_{\lambda`}) 
=\int d^3 r \,\mathbf U_\lambda \cdot \big(  \nabla (\nabla \cdot \mathbf U^*_{\lambda'})-\nabla^2 \mathbf U^*_{\lambda'} \big)$; we use $\nabla \cdot \mathbf U^*_{\lambda'}=0$ and the Helmholtz equation, Equation \eqref{helmotz}, to get $\int d^3 r \, (\nabla \times \mathbf U_\lambda) \cdot (\nabla \times \mathbf U^*_{\lambda`}) = k_\lambda^2  \int d^3 r \,\mathbf U_\lambda \cdot  \mathbf U^*_{\lambda'} = k_\lambda^2 \delta_{\lambda,\lambda'}$. } , with~ $\omega_\lambda = c k_\lambda$. Finally, introducing canonical variables with proper normalization:
\begin{equation}
\iqq_\lambda \equiv \sqrt{\varepsilon_0}\sqrt{\omega_\lambda}\, \il_\lambda \qquad \qquad
\ipp_\lambda \equiv \frac{\sqrt{\varepsilon_0}}{\sqrt{\omega_\lambda}}\, \dot{\il}_\lambda,
\label{change1}
\end{equation}
the total energy of the electromagnetic field $H_{\text{field}}$ can be expressed as
\begin{equation}
H_{\text{field}} =  \sum_{\lambda=1}^{\mathcal{M}} \frac{ \omega_\lambda}{2} \left( \ipp_\lambda^2 + \iqq_\lambda^2 \right),
\label{energy3}
\end{equation}
which has the standard form of $\mathcal{M}$ independent harmonic~oscillators.

\subsubsection{Light--Matter~Interaction}

Expanding the kinetic term in the total Hamiltonian \eqref{H_total}, we get
\begin{equation}
   \frac{1}{2m_e}\left(\mathbf{p} - \qele \mathbf{A}(\mathbf{r},t)\right)^2=   \left(\frac{\mathbf{p}^2}{2 m_e}-  \frac{\qele}{m_e} \mathbf{A}(\mathbf r ,t) \cdot \mathbf{p} 
+ \frac{\qele^2}{2 m_e} \mathbf{A}^2(\mathbf r ,t) \right)
\label{pa}
\end{equation}
The term $- \frac{\qele}{m_e} \mathbf{A}(\mathbf r ,t) \cdot \mathbf{p}$ is the so-called \textit{linear light--matter interaction} or the $\mathbf{p}\cdot\mathbf{A}$ term. The~term \( \frac{\qele^2}{2 m_e} \mathbf{A}^2(\mathbf r ,t) \) is the \textit{diamagnetic term} or quadratic interaction. Using the mode decomposition in \eqref{expansion}, the~total Hamiltonian in \eqref{H_total} reads as
\begin{eqnarray}
H =   \frac{1}{2m_e}  \mathbf{p}^2 +  \qele A_0(\mathbf{r},t)+ \sum_{\lambda=1}^{\mathcal{M}} \frac{ \omega_\lambda}{2} \left( \ipp_\lambda^2 + \iqq_\lambda^2 \right)+H_{\text{int}}
\label{H_total2}
\end{eqnarray}
with the light--matter interaction given by
\begin{eqnarray}
H_{\text{int}}=-\frac{\qele}{m_e} \sum_{\lambda=1}^{\mathcal{M}} \frac{1}{\sqrt{\varepsilon_0} \sqrt{\omega_\lambda}}\iqq_\lambda  \mathbf{U}_\lambda(\mathbf r ) \cdot \mathbf{p}
+
\frac{\qele^2}{2 m_e} \sum_{\lambda=1,\lambda'=1}^{\mathcal{M},\mathcal{M}} 
\frac{1}{\sqrt{\omega_\lambda}\sqrt{\omega_{\lambda'}}\varepsilon_0}\iqq_\lambda \iqq_{\lambda'} \mathbf{U}_\lambda(\mathbf r ) \cdot \mathbf{U}_{\lambda'}(\mathbf r ).
\label{H_int}
\end{eqnarray}

\section{Canonical~Quantization}
\label{sec3}

In most textbooks~\cite{Grynberg2011,Cohen,Scully}, the~standard procedure to move from a classical to a quantum description of a Hamiltonian system is the canonical quantization implemented through the Dirac prescription: the~canonical variables $a$ and $b$ are promoted to quantum operators $\hat a$ and $\hat b$ satisfying commutation relations determined by their classical Poisson brackets\footnote{The classical Poisson bracket, in~general, for~a pair of classical observables $f(a,b)$ and $g(a,b)$, is defined as
\begin{equation}
\{f,g\}=\frac{\partial f}{\partial a}\frac{\partial g}{\partial b}-\frac{\partial f}{\partial b}\frac{\partial g}{\partial a}.
\end{equation}}
and the quantum commutators\footnote{In quantum mechanics, observables are represented by Hermitian operators. The~commutator of two operators $\hat F$ and $\hat G$ is defined as   $[\hat F,\hat G]=\hat F \hat G - \hat G \hat F$.}. In~particular, the~Dirac prescription establishes the correspondence
\begin{equation}
[\hat a,\hat b] = i\hbar \{a,b\}.
\end{equation}
Other observables $A(a,b)$ and $B(a,b)$ are then represented by operators constructed from the canonical operators $A(\hat a,\hat b)$ and $B(\hat a,\hat b)$. This recipe to quantize a system is not free from ambiguities\footnote{For example, the~classical observables satisfy $a\cdot b+b\cdot a=2\; a\cdot b$, but~this equivalence is not true for its quantum version because $\hat a\cdot\hat b$ and $\hat b\cdot\hat a$ are different. This problem will appear in the quantum version of \eqref{pa} with the terms $\mathbf{p}\cdot\mathbf{A}$ and \mbox{$\mathbf{A}\cdot\mathbf{p}$}, but in the Coulomb gauge, this issue will not be problematic.}.

\subsection{Quantization of~Matter}
\label{sec31}

Canonical quantization of matter promotes the classical variables 
$\mathbf r  = (x, y, z)$ and 
$\mathbf p = (p_{x}, p_{y}, p_{z})$ to operators
acting on a Hilbert space. For~example, $x \rightarrow \hat{x}$ and \mbox{$\quad p_{x}\rightarrow \hat{p}_{x}$} so that the Poisson brackets are replaced by commutators according to Dirac's rule \linebreak  $[\hat x, \hat p_{x}] = i\hbar \{x, p_{x}\} = i\hbar$. In~the coordinate (position) representation\footnote{It is a straightforward procedure to show $[\hat x, \hat p] \psi(x) = - x \; i\hbar \frac{d\psi(x)}{dx} + i\hbar \frac{d}{dx} \left( x \; \psi(x) \right) = i\hbar \psi(x)$, where the quantum state in the position representation is $\psi(x)$.}, we get
\begin{equation}
\hat{x} \to x, \qquad\qquad \hat{p} \to -i\hbar \frac{d}{dx}
\end{equation}
For a one-electron system $\mathbf{r}$  in~a rectangular well whose infinite barriers are defined (by the other $i=1, \ldots,\mathcal{N}$ electrons not explicitly simulated)  through the term $V(\mathbf{r}) = eA_0(\mathbf{r},t)$, we get
\begin{eqnarray}
  \left( \frac{1}{2m_e}  \mathbf{p}^2 +  V(\mathbf{r},t)\right) \to   \left(-\frac{\hbar^2}{2m_e}  \nabla^2 +  V(\mathbf{r},t)\right)=  H_{\text{well}}
\label{H_quantumwell}
\end{eqnarray}
with $\nabla=(\partial_{x},\partial_{y},\partial_{z})$. We define the eigenstates of such a single-particle Hamiltonian in the Hilbert space involving only one matter degree of freedom $\mathbf{r}$ as
\begin{eqnarray}
H_{\text{well}} \phi_n (\mathbf{r})= E_{n,e}  \phi_n (\mathbf{r})
\label{matter_eigenstates}
\end{eqnarray}
where $\phi_n (\mathbf{r})$ is the energy eigenstate and $E_{n,e}$ the energy eigenvalue. The~subscript $e$ is to remember that it is an energy from matter (typically electrons). In~principle, there are an infinite number of eigenstates, but~we will consider a finite number
 $n=1, \ldots,N$.

\subsection{Quantization of the Electromagnetic~Field}
\label{sec32}

Canonical quantization of the electromagnetic field follows a procedure 
analogous to the quantization of matter, but~now the dynamical variables 
are field amplitudes instead of particle coordinates. Canonical quantization promotes the classical variables to operators
acting on a Hilbert space $\iqq_\lambda \to \hat \iqq_\lambda$ and $\ipp_\lambda \to \hat \ipp_\lambda$, and~the Poisson brackets are replaced by commutators according to Dirac's 
rule, $[\hat \ipp_\lambda , \hat \iqq_{\lambda'}]= i\hbar \delta_{\lambda,\lambda'}$. In~the coordinate representation of the field, we define these operators as
\begin{equation}
\hat \iqq_\lambda \equiv -i\sqrt{\hbar }\frac{\partial}{\partial \iqqq_\lambda}, \qquad\qquad
\hat \ipp_\lambda \equiv \sqrt{\hbar} \iqqq_\lambda.
\label{change2}
\end{equation}

The quantum Hamiltonian of the electromagnetic field thus becomes
\begin{equation}
\sum_{\lambda=1}^{\mathcal{M}} \frac{ \omega_\lambda}{2} \left( \ipp_\lambda^2 + \iqq_\lambda^2 \right) =  \sum_{\lambda=1}^\mathcal{M} 
\frac{\hbar \omega_\lambda}{2}\left(
-  \frac{\partial^2}{\partial \iqqq_\lambda^2}
+ \iqqq_\lambda^2
\right)=\sum_{\lambda=1}^\mathcal{M} H_{\text{field},\lambda}.
\label{hfield2}
\end{equation}
Each mode $\lambda$ is therefore equivalent to a quantum harmonic oscillator. The~eigenvalue equation is
\begin{equation}
H_{\text{field},\lambda} \; \psi_{m,\lambda}(\iqqq_\lambda)= E_{m,\lambda,p} \; \psi_{m,\lambda}(\iqqq_\lambda)= \hbar \omega_\lambda 
\left( m + \frac{1}{2} \right) \; \psi_{m,\lambda}(\iqqq_\lambda)
\label{light_eigenstates}
\end{equation}
where $\psi_{m,\lambda}(\iqqq_\lambda)$ is the $m$-eigenstate corresponding to the mode $\lambda$ whose eigenenergy  is $E_{m,\lambda,p}$. The index $m = 0,1,2,\dots$ indicates which energy eigenstate we are referring to, but it is also called the ``photon occupation number''.  We are using the subindex $p$ to emphasize that we are dealing with energy linked to~photons. 

\subsection{The Schrödinger~Equation}
\label{sec33}

The translation of classical coordinates into quantum operators for matter, in \mbox{Section~\ref{sec21}}, and~for light, in~Section~\ref{sec22}, maps the classical phase space with coordinates $\mathbf{r}$, $\mathbf{p}$, $\iqq_{\lambda}$, and~$\ipp_{\lambda}$, for $\lambda=1,\ldots,\mathcal{M}$ onto a quantum configuration space\footnote{In quantum mechanics $\mathbf{p} = -i\hbar \nabla$ is a differential operator, so operator ordering matters when it acts on position-dependent functions such as $\mathbf{A}(\mathbf{r})$. Acting on a wavefunction $\psi$, one finds$
(\mathbf{p}\cdot\mathbf{A})\psi = -i\hbar\,\nabla\cdot(\mathbf{A}\psi)
= -i\hbar\left[(\nabla\cdot\mathbf{A})\psi + \mathbf{A}\cdot\nabla\psi\right]
=(\mathbf A\cdot \mathbf p)\psi$, where we have used the Coulomb gauge, $\nabla\cdot\mathbf{A}=0$.
}  described by the coordinates $\mathbf{r}$ and $\iqqq_{\lambda}$  for $\lambda=1,\ldots,\mathcal{M}$. The~quantumness of the light and matter degrees of freedom is then encoded in the wave function $\Psi \equiv \Psi(\mathbf{r},\iqqq_{1},\ldots,\iqqq_{\mathcal M},t)$, which is a solution of the following Schrödinger equation generated by the total Hamiltonian $H$ in \eqref{H_total}:
\begin{align}
i\hbar \frac{\partial \Psi}{\partial t} 
=&  \left(
-\frac{\hbar^2}{2 m_e} \nabla^2 + \qele A_0(\mathbf r,t)
\right) \Psi +
\sum_{\lambda=1}^\mathcal{M} \frac{\hbar \omega_\lambda}{2}\left(
-  \frac{\partial^2}{\partial \iqqq_\lambda^2} + \iqqq_\lambda^2 \right) \Psi
\nonumber\\
& + \frac{\qele}{m_e} \sum_{\lambda=1}^{\mathcal M} \frac{\hbar\sqrt{\hbar} }{\sqrt{\varepsilon_0} \sqrt{\omega_\lambda}} 
\frac{\partial}{\partial \iqqq_\lambda} \mathbf{U}_\lambda(\mathbf r ) \cdot 
\nabla \Psi \nonumber\\
&-  \frac{\qele^2}{2 m_e} \sum_{\lambda=1,\lambda'=1}^{\mathcal M,\mathcal M} \frac{\hbar}{\varepsilon_0 \sqrt{\omega_\lambda} \sqrt{\omega_{\lambda'}}}  
\frac{\partial}{\partial \iqqq_\lambda}\frac{\partial}{\partial \iqqq_{\lambda'}} 
\mathbf{U}_\lambda(\mathbf r ) \cdot \mathbf{U}_{\lambda'}(\mathbf r )\, \Psi
\label{scho_total}
\end{align}
It is very instructive to use the matter eigenstates $\phi_n(\mathbf{r})$ in \eqref{matter_eigenstates}, corresponding in our case to a single electron, and~the light eigenstates $\psi_m(\iqqq_j)$ in \eqref{light_eigenstates}, for each of the $j = 1, \ldots, \mathcal{M}$ electromagnetic modes, to~rewrite the wave function. Let us assume that,  for~each \linebreak  $j$-electromagnetic mode, the~light eigenstates are truncated to $m_j= 1, \ldots, M$:
\begin{equation}
\Psi(\mathbf{r},\iqqq_{1},\ldots,\iqqq_{\mathcal M},t)=\sum_{n,{m_1},...,m_{\mathcal{M}}}^{N,M,...,M}c_{n,m_1,...,m_{\mathcal{M}}}(t)\phi_n(\mathbf{r})\prod_{j=1}^{\mathcal{M}}\psi_{m_j}(\iqqq_{j})
\label{wavefunction}
\end{equation}
where we have defined
\begin{equation}
c_{n,m_1,...,m_{\mathcal{M}}}(t)=\int d\mathbf{r} \int d\iqqq_{1}....\int d\iqqq_{\mathcal M} \phi^*_n(\mathbf{r})\prod_{j=1}^{\mathcal{M}}\psi^*_{m_j}(\iqqq_{j}) \Psi(\mathbf{r},\iqqq_{1},\ldots,\iqqq_{\mathcal M},t)
\label{component}
\end{equation}
whose square value $|c_{n,m_1,...,m_{\mathcal{M}}}(t)|^2$ can be understood as the probability that the light matter system is described by an electron in the eigenstate $\phi_n(\mathbf{r})$ with energy $E_{n,e}$ and the electromagnetic fields with the modes $\prod_{j=1}^{\mathcal{M}}\psi_{m_j}(\iqqq_{j})$ with energy $\sum_{j=1}^{\mathcal{M}} E_{m_j,j,p}$. 

Putting \eqref{wavefunction} into \eqref{scho_total} and using \eqref{matter_eigenstates} and \eqref{light_eigenstates}, and~the orthogonality of the different matter and light eigenstates, we get an equation of motion for the components  $c_{n,m_1,...,m_{\mathcal{M}}}(t)$ in \eqref{component} given by
\begin{eqnarray}
    i\hbar \frac{\partial }{\partial t}c_{n,m_1,...,m_{\mathcal{M}}}(t)&=& (E_{n,e} + \sum_{j=1}^{\mathcal{M}} E_{m_j,j,p})c_{n,m_1,...,m_{\mathcal{M}}}(t) \nonumber\\
    &+& \sum_{n',{m'_1},...,m'_{\mathcal{M}}}^{N,M,...,M}c_{n',m'_1,...,m'_{\mathcal{M}}}(t) \beta_{n,m_1,...,m_{\mathcal{M}};n',m'_1,...,m'_{\mathcal{M}}} \nonumber\\
    &+& \sum_{n',{m'_1},...,m'_{\mathcal{M}}}^{N,M,...,M}c_{n',m'_1,...,m'_{\mathcal{M}}}(t) \gamma_{n,m_1,...,m_{\mathcal{M}};n',m'_1,...,m'_{\mathcal{M}}} \nonumber\\
\label{schoc}
 \end{eqnarray}
with $\beta_{n,m_1,...,m_{\mathcal{M}};n',m'_1,...,m'_{\mathcal{M}}}$ given by
\begin{eqnarray}
\beta_{n,m_1,...,m_{\mathcal{M}};n',m'_1,...,m'_{\mathcal{M}}}=&&\frac{\qele}{m_e} \sum_{\lambda=1}^{\mathcal M} \frac{\hbar\sqrt{\hbar}}{\sqrt{\varepsilon_0} \sqrt{\omega_\lambda}}  \int d\mathbf{r} \phi^*_n(\mathbf{r}) \mathbf{U}_\lambda(\mathbf r ) \cdot \nabla \phi_n(\mathbf{r})\nonumber\\
&& \int d\iqqq_{1}....\int d\iqqq_{\mathcal M} \prod_{j=1}^{\mathcal{M}}\psi^*_{m_j}(\iqqq_{j}) \frac{\partial}{\partial \iqqq_\lambda} \prod_{j'=1}^{\mathcal{M}}\psi_{m_j'}(\iqqq_{j'})
\label{componentbeta}
\end{eqnarray}
and with $\gamma_{n,m_1,...,m_{\mathcal{M}};n',m'_1,...,m'_{\mathcal{M}}}$ given by
\begin{eqnarray}
\gamma_{n,m_1,...,m_{\mathcal{M}};n',m'_1,...,m'_{\mathcal{M}}}=&& -\frac{\qele^2}{2 m_e} \sum_{\lambda=1,\lambda'=1}^{\mathcal M,\mathcal M} \frac{\hbar }{\varepsilon_0 \sqrt{\omega_\lambda} \sqrt{\omega_{\lambda'}}} \int d\mathbf{r}  \phi^*_n(\mathbf{r}) \mathbf{U}_\lambda(\mathbf r ) \cdot \mathbf{A}_{\lambda'}(\mathbf r )  \phi_n(\mathbf{r})\nonumber\\
&& \int d\iqqq_{1}....\int d\iqqq_{\mathcal M}\prod_{j=1}^{\mathcal{M}}\psi^*_{m_j}(\iqqq_{j}) \frac{\partial}{\partial \iqqq_\lambda}\frac{\partial}{\partial \iqqq_{\lambda'}} 
 \prod_{j'=1}^{\mathcal{M}}\psi_{m_j'}(\iqqq_{j'})
\label{componentgamma}
\end{eqnarray}
In fact, the~final expression in \eqref{schoc} can be understood as a matrix equation 
without any explicit reference to the configuration-space degrees of freedom 
$\mathbf{r}, \iqqq_{1}, \ldots, \iqqq_{\mathcal M}$. See Appendix~\ref{appendix} for a discussion of the quantum light without coordinates. Following the notation mentioned in Appendix~\ref{appendix},  we can proceed analogously 
to rewrite the quantum state in \eqref{wavefunction} as
\begin{eqnarray}
\ket{\Psi(t)}= \sum_{n,{m_1},...,m_{\mathcal{M}}}^{N,M,...,M}
c_{n,m_1,...,m_{\mathcal{M}}}(t)
\Big(
\ket{n}_e \otimes 
\ket{m_1}_{\lambda_{1},p} \otimes 
\dots \otimes 
\ket{m_{\mathcal M}}_{\lambda_{\mathcal M},p}
\Big).
\label{quantumstate}
\end{eqnarray}
We have used the bra-ket notation given by $\psi_{m_j}(\iqqq_{j})\equiv \langle \iqqq_{j}| m_j\rangle_{\lambda_j,p}
$.  In~doing so, we lose the ability to visualize the underlying quantum 
phenomena and, more importantly, we remove the possibility of assigning a clear 
ontological meaning to the computations. 

In next ~Section~\ref{sec4}, we will discuss 
how the consideration of the degrees of freedom 
$\mathbf{r}, \iqqq_{1}, \ldots, \iqqq_{\mathcal M}$ 
allows us to provide an ontology for the above formalism. In Sections~\ref{sec5} 
and \ref{sec6}, we will show how this ontology leads to a rather straightforward understanding of what it means to measure photon or electromagnetic properties without invoking words such as \textit{collapse} (with the deep fundamental meaning that the orthodox quantum theory gives to it, pretending to model the whole process of measurement with just a word).

\section{Bohmianization}
\label{sec4}

At this point we already have an ordinary wave function $\Psi(\mathbf{r},\iqqq_{1},\ldots,\iqqq_{\mathcal M},t)$ solution of the ordinary Schrödinger equation, Equation \eqref{scho_total}, that can be used to compute all expectation values relevant to cavity quantum electrodynamics phenomena. Why, then, do we need anything more? The reason is that we are interested in providing an ontology for matter (and light), so that we can meaningfully specify what physically exists and what our theory is about. One may argue that the wave function itself already constitutes the ontology. However, this move does not necessarily improve our sense of ontological clarity. The~wave function lives in a high-dimensional configuration space~\cite{Albert1992}, which makes it difficult to interpret as a concrete physical entity in ordinary three-dimensional space. {Moreover, additional interpretive postulates—such as spontaneous collapse mechanisms, measurement induced collapse or branching structures—are still required to account for the appearance of definite measurement outcomes}. We will see in Section~\ref{sec6} that the Bohmian ontology will provide an straightforward and unambiguous explanation of what it means detecting a photon and its~properties. 

The quantum dynamics determined by the Schrödinger equation, Equation \eqref{scho_total}, in the coordinate space 
$\{\mathbf{r},\iqqq_1,\ldots,\iqqq_{\mathcal{M}}\}$ can be described in terms of quantum trajectories in the same coordinate space. 
The trajectory $\mathbf{r}(t)$ represents the position of the electron as a function of time. 
The trajectory $\iqqq_\lambda(t)$ represents the value of the coordinate associated with the electromagnetic mode $\lambda$ at time $t$. 
By translating $\iqq_{\lambda}(t)$ into $\il_{\lambda}(t)$ through \eqref{change1} and \eqref{change2}, one can recover the quantum electric field as a superposition of well-defined classical field modes $\mathbf{U}_\lambda(\mathbf{r})$ with well-defined time-dependent~amplitudes.

To determine such trajectories, one must identify a continuity equation within the Schrödinger equation, Equation \eqref{scho_total}. 
For this purpose, it is convenient to rewrite the wave function in polar form:
\begin{equation}
\Psi(\mathbf{r},\iqqq_{1},\ldots,\iqqq_{\mathcal M},t)
=
R(\mathbf{r},\iqqq_{1},\ldots,\iqqq_{\mathcal M},t)
\exp{\left({i\frac{S(\mathbf{r},\iqqq_{1},\ldots,\iqqq_{\mathcal M},t)}{\hbar}}\right)}
\label{polar}
\end{equation}
where the modulus 
$R\equiv R(\mathbf{r},\iqqq_{1},\ldots,\iqqq_{\mathcal M},t)$ 
and the phase 
$S \equiv S(\mathbf{r},\iqqq_{1},\ldots,\iqqq_{\mathcal M},t)$ 
are real (meaning not complex) functions. 
Substituting \eqref{polar} into \eqref{scho_total}, a~straightforward calculation yields the following continuity equation \footnote{For matter, the quantum operators of position and momentum are unambiguously defined as $\hat{x} = x$ and $\hat{p} = -i\hbar \partial_x$, respectively, since measurement outcomes (i.e., pointer readings) are ultimately encoded in the position of matter rather than in its velocity. In contrast, for the electromagnetic mode variables, there is an ambiguity in determining which quantum operator should act multiplicatively and which as a derivative, since these variables are not directly measured.
Throughout this paper, we adopt the following convention for the electromagnetic mode variables: $q$ is taken as a multiplicative operator and $s = -i\partial_q$ as a derivative operator. For the numerical simulations, the Bohmian velocities computed in Appendix \ref{ap:trajectories} are derived from the simplified Hamiltonian given in \eqref{eq:QC_SLIM_scho_noap}, where the matter--light interaction (within the dipole approximation) is expressed in terms of the electric field $\mathbf{E}$, which is proportional to $q$ rather than to $s$. The resulting simplification of the numerical implementation is the only argument used to justify our original choice of electromagnetic field operators throughout the paper. The only exception is in the derivation of the continuity equation in \eqref{continuity} and in the Bohmian velocities in this Sec. \ref{sec4}, where we adopt the opposite convention, namely, $s$ as a multiplicative operator and $q = -i\partial_s$ as a derivative operator.
This latter choice is motivated by the fact that the light--matter interaction appears through the vector potential $\mathbf{A}$ in \eqref{scho_total}, which is proportional to $s$ rather than to $q$ as seen in \eqref{expansion}. With this convention, the corresponding expressions are significantly simplified (thus, in the expressions of the Bohmian velocity below, $q$ should be understood as $s$). 
It can be demonstrated that the ambiguity in the choice of operators associated with the electromagnetic mode variables has no net effect on the matter (electron) dynamics.}:
\begin{equation}
\frac{\partial R^2}{\partial t}
+
\nabla \cdot 
\left(
R^2 
\left(
\frac{\nabla S - \qele \mathbf{A}}{m_e}
\right)
\right)
+
\sum_{\lambda=1}^{\mathcal{M}}
\frac{\partial}{\partial \iqqq_\lambda}
\left(
R^2
\frac{\omega_\lambda}{\hbar}
\frac{\partial S}{\partial \iqqq_\lambda}
\right)
=
0
\label{continuity}
\end{equation}

Equation \eqref{continuity} allows us to define the probability density
\begin{equation}
\rho
\equiv
R^2
=
|\Psi(\mathbf{r},\iqqq_{1},\ldots,\iqqq_{\mathcal M},t)|^2
\end{equation}
and a quantum current density for the electron, $\mathbf{J}_e$, related to the electron velocity as
\begin{equation}
\mathbf{v}_e
=
\frac{\mathbf{J}_e}{\rho}
=
\frac{\nabla S - \qele \mathbf{A}}{m_e}
\label{v_e}
\end{equation}
as well as a quantum current density for the electromagnetic mode coordinates, $J_\lambda$, related to their velocities by
\begin{equation}
v_\lambda
=
\frac{J_\lambda}{\rho}
=
\frac{\omega_\lambda}{\hbar}
\frac{\partial S}{\partial \iqqq_\lambda}
\label{v_l}
\end{equation}

Finally, the~continuity equation, Equation \eqref{continuity}, can be rewritten as follows\footnote{It is important to realize that this continuity equation in \eqref{continuityensemble} refers to a set of trajectories, each corresponding to a different experiment. In~contrast, the~continuity equation in \eqref{continuitysingle} referred to a set of $\mathcal{N}$ trajectories within a single experiment. The~fact that trajectories associated with experiments that are not actually performed still influence a given experiment is at the heart of quantum theory.}.
\begin{equation}
\frac{\partial \rho}{\partial t}
+
\nabla \cdot \left( \rho \, \mathbf{v}_e \right)
+
\sum_{\lambda=1}^{\mathcal{M}}
\frac{\partial}{\partial \iqqq_\lambda}
\left( \rho \, v_\lambda \right)
=
0
\label{continuityensemble}
\end{equation}
The above equation is a differential form of the $|\Psi|$-equivariance of the trajectories: if the initial $t=0$ distribution of trajectories  satisfies $
|\Psi(\mathbf{r},\iqqq_{1},\ldots,\iqqq_{\mathcal M},0)|^2$, then under~the velocity fields defined in \eqref{v_e} and \eqref{v_l}, the~previous set of trajectories satisfies $
|\Psi(\mathbf{r},\iqqq_{1},\ldots,\iqqq_{\mathcal M},t)|^2$ for all times. 
In other words, the~probability density transported by the trajectory dynamics evolves consistently with the Schrödinger evolution of the wave function, ensuring preservation of the Born~rule.

Disregarding a bit the mathematical rigor\footnote{The delta function can only be correctly interpreted when integrated over a region of configuration space.}, the $|\Psi|$-equivariance implicit in \eqref{continuityensemble} can be rewritten as
\begin{equation}
|\Psi(\mathbf{r},\iqqq_{1},\ldots,\iqqq_{\mathcal M},t)|^2
= \frac{1}{N_\text{exp}} 
\sum_{\eta=1}^{N_{\text{exp}}} 
\delta(\mathbf{r}-\mathbf{r}^{(\eta)}(t)) 
\delta(q_1-q_1^{(\eta)}(t)) \cdots \delta(q_\mathcal{M}-q_\mathcal{M}^{(\eta)}(t)) ,
\label{equivariance}
\end{equation}
where $\mathbf{r}^{(\eta)}(t)$ is one particular position of the matter particle and, similarly, $q_1^{(\eta)}(t), \ldots, q_\mathcal{M}^{(\eta)}(t)$ is one particular set of coefficients that define the electromagnetic field through the expression \eqref{expansion}. From~\eqref{equivariance}, it is evident that a quantum electron can be interpreted as a superposition of electrons with well-defined positions and, likewise, a~quantum field can be interpreted as a superposition of well-defined electromagnetic fields. The~parameter $N_{\text{exp}}$ is the number of experiments done in a laboratory for a quantum system prepared with the initial wave function $\Psi(\mathbf{r},\iqqq_{1},\ldots,\iqqq_{\mathcal M},0)$. Strictly speaking, we need $N_{\text{exp}} \to \infty$ to satisfy the identity in equations  \eqref{equivariance} between trajectories and probability presence. 

\section{Born Rule and Statistical~Predictions}
\label{sec5}

The expression of equivariance in \eqref{equivariance} is the fundamental ingredient ensuring that the empirical results obtained by considering only the wave function $\Psi(\mathbf{r},\iqqq_{1},\ldots,\iqqq_{\mathcal M},t)$ and those obtained by considering (the wave function together with) a set of Bohmian trajectories $\mathbf{r}^{(\eta)}(t),q_1^{(\eta)}(t),\dots,q_\mathcal{M}^{(\eta)}(t)$ with $\eta=1,\ldots,N_{\text{exp}}$ are fully~equivalent.

A central advantage of the Bohmian account of quantum electrodynamics (and quantum mechanics in general) is that it provides a natural explanation of the measurement process without invoking the word \textit{collapse}, with the deep fundamental meaning that the orthodox quantum theory gives to such word. In~orthodox quantum theory, \textit{collapse} is introduced as an additional postulate to brake the quantum superposition in order to recover the Born rule. By~contrast, within~Bohmian mechanics, the~Born rule is not an independent axiom: it follows from equivariance together with the quantum equilibrium~hypothesis.

The first step of the measurement formalism (channelization) is an explicit realization of the von Neumann measurement model. A~quantum system is coupled to a pointer in such a way that information about an observable is transferred to the pointer degree of freedom $y$ through a unitary~interaction.

We write the initial state in \eqref{quantumstate} at $t=0$ as
\begin{equation}
\ket{\Psi(0)}=\sum_s c_s(0)\ket{s},
\end{equation}
where $s$ is a unique index that identifies $(n,m_1,\dots,m_{\mathcal M})$ and $\ket{s}=\ket{n}_e\otimes\ket{m_1}_{\lambda_1,p}\otimes\dots\otimes
\ket{m_{\mathcal M}}_{\lambda_{\mathcal M},p}$ in the previous notation. The~measurement interaction is modeled by a von Neumann-type Hamiltonian
\begin{equation}
\hat H_{\text{int}}=\mu\,\hat S\otimes\hat P_{y},
\end{equation}
where $\hat S\ket{s}=s\ket{s}$, $\hat P_{y}$ is the momentum operator of the pointer $y$, and~$\mu$ is the coupling constant. After~a short interaction time $\Delta t$, the~evolution operator is
\begin{equation}
\hat {\mathbb U}=\exp\!\left(-\frac{i}{\hbar}\mu\Delta t\,\hat S\otimes\hat P_{y}\right).
\end{equation}
Defining $\ket{\varphi}$ as the state of the pointer, the~operator $\hat {\mathbb U}$ acts on the product state \linebreak  $\ket{\Phi(0)}=\ket{\Psi(0)}\otimes\ket{\varphi(0)}$ as
\begin{equation}
\ket{\Phi(t)}=\sum_s c_s(t)\ket{s}\otimes
\exp\!\left(-\frac{i}{\hbar}\mu\Delta t\, s\,\hat P_{y}\right)\ket{\varphi(0)}.
\end{equation}
Since $\hat P_{y}$ generates spatial translations on the coordinate $y$, we obtain
\begin{equation}
\ket{\Phi(t)}=\sum_s c_s(t)\ket{s}\otimes\ket{\varphi_s(t)},
\end{equation}
with $|\varphi_s(t)\rangle$ a shifted pointer (ket) state. In the position representation, we get%Please check intended meaning has been retained
\begin{equation}
\langle y|\varphi_s(t)\rangle=\varphi(y-y_s,t).
\end{equation}
where we have defined $y_s=\mu\Delta t\, s$. In~coordinate representation, with~$\xi$ denoting collectively the system coordinates $\bra{\xi} \equiv \bra{\mathbf{r}}\otimes \bra{\iqqq_{1}}\otimes...\otimes\bra{\iqqq_{\mathcal M}}$, the~total wave function reads as
\begin{equation}
\Phi(\xi,y,t)=\sum_s c_s(t)\,\Psi_s(\xi)\,\varphi(y-y_s,t),
\label{channel}
\end{equation}
where $\Psi_s(\xi)=\langle \xi|s\rangle$. If~one ignores the Bohmian configuration (i.e. the well defined trajectories), the~superposition in \eqref{channel} must be broken by postulating collapse. Assuming the pointer packets have disjoint supports $Y_s$, the process of \textit{collapse} selects one branch $s_0$ (How? ``shut up and calculate''), yielding
\begin{equation}
\Phi(\xi,y,t)\approx c_{s_0}(t)\Psi_{s_0}(\xi)\varphi(y-y_{s_0},t).
\label{conditional}
\end{equation}
The probability of outcome $s_0$ is then
\begin{equation}
P(s_0,t)=\int_{Y_{s_0}} dy \int d\xi\,|\Phi(\xi,y,t)|^2
=|c_{s_0}(t)|^2,
\label{result1}
\end{equation}
which is the Born rule. However, because~of the superposition principle, the~unitary evolutions of the Schrödinger equation cannot transform \eqref{channel} into a single branch in \eqref{conditional}; collapse must therefore be added as an extra postulate in the orthodox explanation of~measurement.

In Bohmian mechanics, the~trajectories $\xi(t),y(t)$ evolve under the guidance equation generated by $\Phi(\xi,y,t)$. Once the wave packets in \eqref{channel} become disjoint\footnote{In realistic experimental scenarios, exact disjointness is not required; it suffices that the supports be approximately disjoint, which guarantees an adequate level of distinguishability between the pointer states.}, the~actual pointer position $y(t)$ lies in one and only one region $Y_{s_0}$. The~measurement outcome is simply the value indicated by this~configuration.

The other branches do not influence the actual configuration\footnote{The other branches are empty wave functions in the sense that there is no Bohmian trajectory on their supports. As~long as the global evolution remains under the Markovian regime---notice that a realistic macroscopic pointer deals with $10^{23}$ degrees of freedom---we can assume that such an empty wave function in the $10^{23}$-dimensional configuration space will never interfere again with the non-empty wave function \eqref{conditional}.}. In~this sense, wave function collapse emerges effectively at the level of the \textit{conditional wave function of the subsystem}~\cite{TravisNorsen_cond_wf},
\begin{equation}
\Psi^\eta(\xi,t)=\Phi(\xi,y^\eta(t),t),
\label{wavemeas}
\end{equation}
without modifying the Schr\"odinger dynamics. We have used the superscript $\eta$ to remind that many trajectories $y^\eta(t)$ are possible, as~in \eqref{equivariance}, each one corresponding to a different experiment $\eta=1,\ldots, N_{\text{exp}}$. Each experiment giving different outcomes of the~measurement. 

Since the Bohmian trajectories are fully determined by the wave function, and~the wave function itself is fully determined by the Schr\"odinger equation, one may wonder why the experimental results remain unpredictable in the laboratory. The~reason is that the preparation of an experiment is itself another physical experiment, and~one cannot prepare initial conditions with a precision better\footnote{In principle, one could attempt to prepare an initial wave function sharply peaked around a configuration, 
$|\Phi(\xi,y,0)|^2 \approx \delta(\xi-a)\,\delta(y-b)$. 
In that case, all preparations would satisfy $\xi^{(\eta)}(0)\approx a$ and $y^{(\eta)}(0)\approx b$. 
However, such $\delta$-wave functions are not prepared as initial quantum states in laboratories since they are ill-defined because they require arbitrarily large energies.} than the quantum equilibrium distribution $|\Phi(\xi,y,0)|^2$, see~\cite{Durr1992}. Thus, Bohmian trajectories are deterministic at the ontological level, yet unpredictable at the empirical~level.

If we repeat the same experiment $N_{\text{exp}}$ times with the same preparation of the initial wave function $\Phi(\xi,y,0)$, this implies a random set of initial trajectories $\xi^{(\eta)}(0),y^{(\eta)}(0)$ selected according to the distribution $|\Phi(\xi,y,0)|^2$, with~$\eta=1,\ldots,N_{\text{exp}}$. We may then ask what is the frequentist probability of measuring the value $s_0$. The~answer is
\begin{eqnarray}
P(s_0,t)=\frac{N_{s_0}}{N_{\text{exp}}}
&=&\frac{1}{N_{\text{exp}}}\sum_{\eta=1}^{N_{\text{exp}}}
\int_{Y_{s_0}} dy \; \delta(y-y^{(\eta)}(t)) \nonumber\\
&=&\frac{1}{N_{\text{exp}}}\sum_{\eta=1}^{N_{\text{exp}}}
\int_{Y_{s_0}} dy \; \delta(y-y^{(\eta)}(t))
\int_{-\infty}^{\infty} d\xi\,
\delta(\xi-\xi^{(\eta)}(t)) \nonumber\\
&=&\int_{Y_{s_0}} dy \int_{-\infty}^{\infty} d\xi \,
|\Phi(\xi,y,t)|^2
= |c_{s_0}(t)|^2,
\label{result2}
\end{eqnarray}
where we have  use the identity $\int_{-\infty}^{\infty} d\xi \, \delta(\xi-\xi^{(\eta)}(t))=1$ and the equivariance relation from \eqref{equivariance}, including the pointer degree of freedom, written here as
\begin{equation}
|\Phi(\xi,y,t)|^2=
\frac{1}{N_{\text{exp}}}
\sum_{\eta=1}^{N_{\text{exp}}}
\delta(\xi-\xi^{(\eta)}(t))
\delta(y-y^{(\eta)}(t)),
\end{equation}
Of course, by~construction, \eqref{result1} and \eqref{result2} are identical. Thus, in~Bohmian mechanics, the~Born rule is not a separate dynamical axiom but a consequence of unitary Schr\"odinger evolution, the~guidance equation, and~the equivariance of the $|\Phi|^2$ distribution (under the assumption of the quantum equilibrium hypothesis~\cite{Durr1992}).

\section{Photon Partition~Noise}
\label{sec6}

The formalism introduced in the previous sections is applied here to show how photon partition noise arises without the need to assume an ontic nature for photons. First, a~model of two spatially separated electrons interacting with light, but~without measuring apparatus, is presented to gain insight into the dynamics of light--matter interaction in general, and~Rabi oscillations in particular. Then, a~model incorporating two measuring apparatuses, each coupled to one electron, is discussed. This model shows how photon detection occurs either in one detector or in the other, but~not in both. In~our formalism, the~``detected'' partition noise of the photons is, in~fact, an~artifact of the particle nature of the matter particles in the detectors. Consequently, there is no need to attribute any ontic character to~photons.

\subsection{Simulation of a Non-Measured Light--Matter~System}
 
We consider a quantum--optical setup depicted in Figure~\ref{fig:set_up_non_measured}a, consisting of an optical cavity of length $L_c$ (related to a cavity fundamental angular frequency $\omega_c$) containing two spatially separated quantum wells, each hosting a single electron. We consider only one electromagnetic mode of the cavity, whose fundamental frequency is tuned to resonate with the transition of the two-level electronic systems in the quantum wells\footnote{Many more particles are involved in this experiment (in particular those of the matter particles required to create the optical cavity and the quantum wells) that are not included in the simulation. Following the discussion in the introduction, one can argue that the interaction of the other non-simulated particles with the two simulated electrons is effectively incorporated through the electromagnetic field $\mathbf{E}$ and $\mathbf{B}$, and~the potential barriers of the quantum wells and optical~cavity.}.

\vspace{-3pt}
 \begin{figure}[h!]
 %    \centering
  \hspace{-0.18cm}   \includegraphics[width=0.9\linewidth]{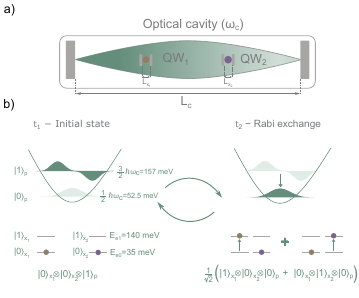}
     \caption{
(\textbf{a}) Sketch of the simulated system. An~optical cavity containing a single electromagnetic mode of frequency $\omega_c$, with~two quantum wells located inside the cavity. 
(\textbf{b}) Two-level energy diagram of the different components of the system at two instants of the Rabi oscillations. 
The initial state at time $t_1$, $\ket{001}$, corresponds to the electromagnetic field in the excited state while both electrons are in their ground states. 
At time $t_2$, the~excitation is coherently transferred to the electronic subsystem, producing the superposition $\frac{1}{\sqrt{2}}\left(\ket{100} + \ket{010}\right)$. 
The excitation is subsequently exchanged periodically between the photonic and electronic degrees of freedom through Rabi oscillations.
}
     \label{fig:set_up_non_measured}
 \end{figure}

\subsubsection{Model  Without Measuring Apparatus
}
\label{sec611}

The one-mode electromagnetic field inside the cavity is quantized, and~we consider the two lowest energy levels of the electromagnetic mode. Identically, the electrons in the quantum wells are quantized and we consider the two lowest energy levels for the electrons. Due to the resonant coupling, energy is transferred from the electromagnetic field to the electronic subsystem and vice~versa through Rabi oscillations (Figure~\ref{fig:set_up_non_measured}b). To~simplify the notation, sometimes we are using the ket notation; for example, $\ket{001}= \ket{0}_{x_1}\otimes\ket{0}_{x_2}\otimes\ket{1}_{q}$ to refer to the wave function $\phi_0(x_1)\phi_0(x_2)\psi_1(q)$. The~Schrödinger equation describing the interaction among one electron and a single electromagnetic field was derived in Section~\ref{sec33}. Some simplifications are needed to arrive at a computationally feasible scenario. First, we assume that the two electrons are sufficiently far apart so that their Coulomb interaction can be neglected. Therefore, the~scalar potential can be assumed to include only the contributions from the two quantum wells due to other, non-simulated particles $A_0(\mathbf r_1, \mathbf r_2,t)= V(\mathbf r_1) + V(\mathbf r_2)$. A~typical additional approximation in these scenarios is the so-called long-wavelength approximation, based on the assumption that the cavity length satisfies $L_c \gg L_{x_1}, L_{x_2}$, where $L_{x_1}$ and $L_{x_2}$ denote the spatial extensions of the electronic quantum wells. Under~this condition, the~electromagnetic field varies negligibly over the electronic region and the spatial dependence of the vector potential can be neglected. For~simplicity, we also assume that each electron is described by only one degree of freedom, $x_1$ for the first electron and $x_2$ for the second. Under~these simplifications one arrives at
\begin{align}
i\hbar \frac{\partial \Psi}{\partial t} 
=&  \left(
-\frac{\hbar^2}{2 m_e} \frac{\partial^2}{\partial x_1^2} + V(x_1)  -\frac{\hbar^2}{2 m_e} \frac{\partial^2}{\partial x_2^2}+ V(x_2)
\right) \Psi +
 \frac{\hbar \omega_c}{2}\left(
-  \frac{\partial^2}{\partial \iqqq^2} + \iqqq^2 \right) \Psi
\nonumber\\
& + \frac{\qele}{ m_e}  \frac{\hbar\sqrt{\hbar} U_0 }{\sqrt{\varepsilon_0} \sqrt{\omega_c}} 
\frac{\partial}{\partial \iqqq} 
\frac{\partial}{\partial x_1} \Psi  + \frac{\qele}{ m_e}  \frac{\hbar\sqrt{\hbar} U_0}{\sqrt{\varepsilon_0} \sqrt{\omega_c}} 
\frac{\partial}{\partial \iqqq} 
\frac{\partial}{\partial x_2} \Psi - \frac{\qele^2}{2 m_e} \frac{\hbar}{\varepsilon_0 \omega_c}  
\frac{\partial^2}{\partial \iqqq^2}
U_0^2\, \Psi,
\label{scho_total_aprox}
\end{align}
where $U_0$ is the spatial part of the potential vector, which has been assumed to be uniform in the spatial regions where electrons moves due to the long-wavelength approximation.
One final step is to perform a gauge transformation to write the Hamiltonian in dipolar form (see Appendix~\ref{ap:no_meas} for a full derivation).  

After all these steps, one arrives at the following simplified Schrödinger equation dealing with the configuration variables $\xi=(x_1,x_2,q)$, representing two electrons and one electromagnetic mode:
\begin{eqnarray}
    i\hbar \frac{\partial\Psi}{\partial t}&=&\left( -\frac{\hbar^2}{2m_e}\frac{\partial^2}{\partial x_1^2} + V(x_1)\right)\Psi +\left(-\frac{\hbar^2}{2m_e}\frac{\partial^2}{\partial x_2^2} + V(x_2)\right)\Psi   
\nonumber\\&+&\frac{\hbar\omega_c}{2}\left(-\frac{\partial^2}{\partial q^2} + q^2\right)\Psi  + \alpha q (x_1+x_2)\Psi,
    \label{eq:QC_SLIM_scho_noap}
\end{eqnarray}
where we have defined the total wave function as $\Psi \equiv \Psi(x_1,x_2,q,t)$ and the parameter that controls the coupling between the light and the matter as $\alpha= \frac{\qele\sqrt{\hbar \omega_c }U_0}{\sqrt{\varepsilon_0}}$.  To~solve the Schrödinger equation, Equation \eqref{eq:QC_SLIM_scho_noap}, we expand the total wave function in a truncated product basis,
\begin{equation}
\Psi(x_1,x_2,q,t)= \sum_{n=0}^{1}\sum_{m=0}^{1}\sum_{k=0}^{1}
c_{nmk}(t)\phi_n(x_1)\phi_m(x_2)\psi_k(q),
\label{trunwave}
\end{equation} 
where $\phi_0(x_1)$ and $\phi_1(x_1)$ are the first two eigenstates for the electron $x_1$ in the quantum well defined from \eqref{matter_eigenstates}. Equivalently, we define $\phi_0(x_2)$ and $\phi_1(x_2)$ for the second electron $x_2$ in the other quantum well. Finally, the~cavity mode is described by the first two eigenfunctions $\psi_0(q)$ and $\psi_1(q)$ of the harmonic oscillator defined in \eqref{light_eigenstates}. 

Substituting the expansion \eqref{trunwave} into \eqref{eq:QC_SLIM_scho_noap} and projecting onto the basis states, in~agreement with the development in Section~\ref{sec33}, yields the coupled equations of motion for the coefficients $c_{nmk}(t)$:
\begin{align}
i\hbar \frac{d}{dt} c_{nmk}(t)=
E_{nmk}\, c_{nmk}(t)
+\sum_{n'=0}^{1}\sum_{k'=0}^{1} 
\beta^{x_1}_{nk,n'k'}\,c_{n'mk'}(t)
+\sum_{m'=0}^{1}\sum_{k'=0}^{1} 
\beta^{x_2}_{mk,m'k'}\,c_{nm'k'}(t)
\label{eq:c_nmk},
\end{align}
where the energy $E_{nmk}=E_{n,e}+E_{m,e}+E_{k,p}$ is defined. We also introduce the matrix element
\begin{equation}
\beta^{x_1}_{nk,n'k'}=
\alpha
\int \phi_n^*(x_1)\, x_1 \,\phi_{n'}(x_1)\, dx_1
\int \psi^*_k(q)\, q \,\psi_{k'}(q)\, dq
\end{equation}
whose value is zero whenever  $n=n'$, since 
$\int \phi^*_n(x_1)\, x_1 \,\phi_{n}(x_1)\, dx_1=0$, or~when $k=k'$, since 
$\int \psi_k^*(q)\, q \,\psi_{k}(q)\, dq=0$.  Similarly, for~the matrix element
\begin{equation}
\beta^{x_2}_{mk,m'k'}=
\alpha
\int \phi^*_m(x_2)\, x_2 \,\phi_{m'}(x_2)\, dx_2
\int \psi^*_k(q)\, q \,\psi_{k'}(q)\, dq,
\end{equation}
a zero value is obtained when $m=m'$, since 
$\int \phi^*_m(x_2) x_2 \phi_{m}(x_2)\, dx_2=0$, or~when $k=k$, since 
$\int \psi^*_k(q)\, q \,\psi_{k}(q)\, dq=0$.

This structure of the interaction terms in \eqref{eq:c_nmk} implies that the Hamiltonian only connects basis states $\phi_n(x_1)\phi_m(x_2)\psi_k(q)$ with $\phi_{n'}(x_1)\phi_{m'}(x_2)\psi_{k'}(q)$ when two subscripts change (from $0\to1$ or $1\to0$). Therefore, starting from a state with an even number of excitations ($1$s subscripts), the~dynamics will always remain within the sub-space with an even number of excitations (even number of $1$s in the three subscripts). The~same holds for states with an odd number of excitations. Consequently, the~total Hilbert space splits as a direct sum $\mathcal{H}=\mathcal{H}_{\mathrm{even}} \oplus \mathcal{H}_{\mathrm{odd}}$ with
\begin{equation}
\mathcal{H}_{\mathrm{even}}
=
\mathrm{span}
\{ {\ket{000},\ket{110},\ket{101},\ket{011}} \},
\label{up}
\end{equation}
and
\begin{equation}
\mathcal{H}_{\mathrm{odd}}
=
\mathrm{span}
\{\ket{100},\ket{010},\ket{001},\ket{111}\}.
\end{equation}
We remind the use of the ket notation here where, for~example, $\ket{001}= \ket{0}_{x_1}\otimes\ket{0}_{x_2}\otimes\ket{1}_{q}$ refers to the wave function $\phi_0(x_1)\phi_0(x_2)\psi_1(q)$.

\subsubsection{Numerical~Results  Without Measuring Apparatus
}

In our simulation we consider a cavity length equal to $L_c$ = 5908.4 nm, corresponding to a cavity angular frequency 
$\omega_c = 159.4\,\mathrm{THz}$. The~electronic level spacing of each quantum well is chosen to satisfy the resonance condition $E_{1,e} - E_{0,e} = \hbar\omega_c=0.105 \; \text{eV}$. It is well-known that without such resonant conditions light and matter do not interact~\cite{Grynberg2011}. This resonant condition fixes the length of the quantum well to
$L_{x_1}=L_{x_2} = 16\,\mathrm{nm}$ when using an effective electron mass $m_e=0.042m_{0}$, where $m_{0}$ the free electron mass\footnote{The concept of effective electron mass allows electrons in the semiconductor forming the quantum well to be treated as quasi-free particles, incorporating the influence of the periodic lattice without explicitly modeling the atomic structure. The~value $m_e=0.042m_{0}$ is characteristic of Indium Gallium Arsenide, a~material widely used in quantum well structures for high-speed electronic and photonic applications.}. The~light--matter coupling constant is set to $\alpha = 6.24 \times 10^{-3}\,\mathrm{eV/nm}$. For~$N=2$ symmetrically coupled electrons, the~Rabi frequency is given by $\Omega_R = \sqrt{N}\,\frac{\alpha}{\hbar}\,\beta_{nk,n'k'}= 54.6\,\mathrm{THz}$. This implies a Rabi period of $T_R \sim 115\,\mathrm{fs}$.

Figure~\ref{fig:landa_vs_things_0} shows the numerical solution of \eqref{eq:c_nmk} over four
Rabi periods $T_R$. The~energy of the system is initially assigned to the photon excited state $\ket{001}$, with~both electrons in the ground state, by~fixing $|c_{001}(0)|^2=1$ and setting all other coefficients to zero (see Figure~\ref{fig:set_up_non_measured}b, left). The~initial state $\ket{001}$ belongs to $\mathcal{H}_{\mathrm{odd}}$. Therefore, the~unitary dynamics discussed in \eqref{up} guarantee that the state remains in $\mathcal{H}_{\mathrm{odd}}$ for all times, and~no amplitude is ever transferred to $\mathcal{H}_{\mathrm{even}}$. For~this reason, the~numerical results will only show states restricted to $\mathcal{H}_{\mathrm{odd}}$.

From a naive picture, one could argue that 
initial energy stored in the cavity mode $E_{1,p}$ in $\ket{001}$ is transferred
to the energy $E_{1,e}$ of the first electron in $\ket{100}$. Notice that
$E_{1,p}-E_{0,p}=\hbar \omega_c=E_{1,e}-E_{0,e}$. Due to the symmetry of the
Hamiltonian with respect to electrons 1 and 2, it is also possible that the
initial energy stored in the cavity mode in $\ket{001}$ is transferred
to the energy $E_{1,e}$ of the second electron in $\ket{010}$. In~fact, as~shown
in Figure~\ref{fig:landa_vs_things_0} and schematically depicted in
Figure~\ref{fig:set_up_non_measured}b (right), after~half a Rabi period the
system evolves into the symmetric superposition $\frac{1}{\sqrt{2}}\left(\ket{100}+\ket{010}\right)$, indicating that the initial (expectation value) of the energy of the electromagnetic field (what we call
the photon $\hbar \omega_c$) is divided into two and shared between both
electrons, i.e., \mbox{$\langle H_{x1}(T_R/2))\rangle=\langle H_{x2}(T_R/2))\rangle=0.087$ eV}  when 
$|c_{100}(T_R/2)|^2=|c_{010}(T_R/2)|^2=0.5$. Of~course, this scenario also
ensures conservation of (the expectation) value of the energy\footnote{Notice that the energy associated with
the interaction terms in the Hamiltonian is almost negligible due to the small
value of $\alpha$ in the strong light--matter coupling regime considered in
this work. For~larger values of $\alpha$, corresponding to the ultra-strong
coupling regime, the~situation could be quite different.}, as~depicted in
Figure~\ref{fig:landa_vs_things_0}. Unitarity is preserved throughout the
evolution, as~shown by the conservation of the total~probability.

What may seem puzzling (or at least in disagreement with photon partition
noise experiments) is that the initial energy of the photon $\hbar \omega_c=0.105$ eV
is distributed between the energy of electron~1 at positions $x_1$  given by \mbox{$\langle H_{x1}(T_R/2))\rangle=0.035+0.105/2=0.087$} eV, {and~the energy of electron~2 at position $x_2$ given by $\langle H_{x2}(T_R/2))\rangle=0.035+0.105/2=0.087$ eV}. But~is not $\hbar \omega_c=0.105$ eV a quantized
amount of energy that must be detected as a whole? Another way of asking this
question is the following. Why do we not observe in
Figure~\ref{fig:landa_vs_things_0} a process in which the energy of the
electromagnetic field is transferred only to the first electron without 
breaking the quantum of energy $\hbar \omega_c=0.105$ eV  (i.e., the~photon)? In terms of
wave functions, this non-breaking-photon evolution would correspond to
\begin{eqnarray}
&&\phi_0(x_1)\phi_0(x_2)\psi_1(q) \to \phi_1(x_1)\phi_0(x_2)\psi_0(q) \nonumber\\
&&\ket{001} \qquad \qquad \qquad \to \ket{100}
\label{sup1}
\end{eqnarray}
where the energy of the electron~1 at the position $x_1$ (when the other electron remains {with its initial ground state energy) would be given by \mbox{$\langle H_{x1}(T_R/2))\rangle=0.035+0.105=0.140$ eV}}.
Of course, another way of transferring energy between light and matter
without breaking the quantum of energy $\hbar \omega_c=0.105$ eV (i.e., the~photon) at any
time would be obtained by interchanging the roles of the electrons in the
previous non-breaking-photon \mbox{evolution~process}:
\begin{eqnarray}
&&\phi_0(x_1)\phi_0(x_2)\psi_1(q) \to \phi_0(x_1)\phi_1(x_2)\psi_0(q) \nonumber\\
&&\ket{001} \qquad \qquad \qquad \to \ket{010}\qquad \qquad \qquad 
\label{sup2}
\end{eqnarray}
Now, the~energy of the second electron 
at the position $x_2$ (when the other electron remains with its initial ground state energy) would be given by the electron ground state plus the photon energy, \mbox{$\langle H_{x2}(T_R/2))\rangle=0.035+0.105=0.140$ eV}
.  However, what we observe in the simulation is simply the superposition of both non-breaking-photon evolutions given by \eqref{sup1} and \eqref{sup2}, analogous to the superposition of a dead and an alive cat. But, such a superposition seems to indicate that half of the photon energy has moved to  $x_1$ and the other half to  $x_2$. Are the two non-breaking-photon evolutions in \eqref{sup1} and \eqref{sup2} themselves real, or~is the superposition the only real entity? Here, the~word ``real'' refers to ontic elements. In~orthodox quantum mechanics, reality appears when the properties of the system are measured; strictly speaking, we have measured neither \eqref{sup1} nor \eqref{sup2}, nor the~superposition.

In our Bohmian approach, on~the contrary, we do have ontic variables: the positions $x_1(t)$ and $x_2(t)$ of the electrons and the parameter $q(t)$ associated with the light, at~any time, even if the light--matter system has not yet been measured. At~this point, one might be tempted to say that different Bohmian trajectories would produce the above non-breaking-photon evolutions. {{However, this is not true. In~the Bohmian picture, the~only (primitive) ontic elements are the particle positions $x_1(t)$ and $x_2(t)$ at all times, while quantities such as velocities and energies depend on the trajectory and also on the wave function (i.e., they are functionals of the wave function). In~particular, knowledge of the Bohmian trajectory $x_1(t)$ alone is not sufficient to determine a definite value of the electronic energy associated with electron~1; the global wave function must also be known. If~the wave function is a superposition of different energy eigenstates, it can be quite difficult to anticipate the value of the energy associated with electron~1 by knowing only $x_1(t)$. On~the contrary, if~the global wave function is constructed as a sum of different branches, each with dynamically disjoint support and each associated with a single energy eigenstate, then the energy can be readily inferred from the particle position $x_1(t)$ (by identifying on which support of the energy eigenstates the particle lies). In~our particular non-measurement simulation case, as~long as the supports of $\phi_0(x_1)\phi_1(x_2)\psi_0(q)$ and $\phi_1(x_1)\phi_0(x_2)\psi_0(q)$ overlap in configuration space, the~Bohmian configuration is guided by both components. Because~of this overlap, the~energy associated with the trajectory $x_1(t)$ can take rather unexpected values—for example, the energy linked to the first electron %Please check intended meaning has been retained
 in Figure~\ref{fig:landa_vs_things_0}, $\langle H_{x1}(T_R/2))\rangle=0.035+0.105/2=0.087$ eV,  is much lower than the sum of the electron ground-state energy and the photon energy, $0.035 + 0.105 = 0.140~\text{eV}$; the same happens to the second electron %Please check intended meaning has been retained
 —and hence no definite ``photon absorption event'' involving only one electron can appear prior to measurement.}

The apparent paradox that our (orthodox and Bohmian) results display a breaking-photon evolution---each electron ``absorbs%Please check intended meaning has been retained
'' half of the photon energy---while experiments in the laboratory show non-breaking-photon events---one electron ``absorbs%Please check intended meaning has been retained
'' the whole of the photon energy---is, as~anticipated, easily resolved: the light--matter system considered above has not yet been measured. In~other words, we are comparing ``apples and oranges''. One of the most common sources of misunderstanding in quantum mechanics is the assumption that the properties of an unmeasured system are the same as those of a measured one. This is not the case. Even two different measurement setups may lead to different evolutions, which is why any quantum theory consistent with experiments must be contextual. It is one of the great merits of Bohmian theory (as discussed in Section~\ref{sec5}) to show explicitly why this happens: measuring a system means coupling another system to the original one, and~two coupled systems behave differently from a single isolated~system. 

\vspace{-6pt}
\begin{figure}[h!]
  %  \centering
\includegraphics[width=1\linewidth]{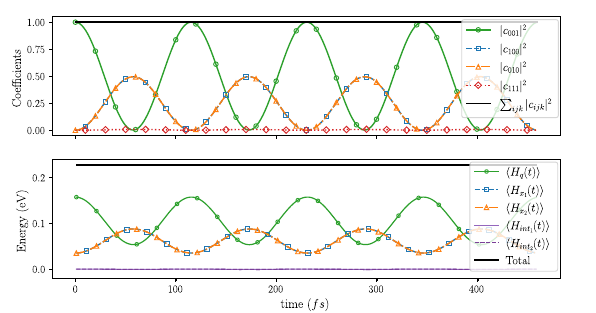}
    \caption{Unitary dynamics of the non-measured system under resonant coherent coupling. 
\linebreak  \textbf{Top panel}: time evolution of the populations $|c_{nmk}(t)|^2$ within the odd-parity sub-space over four Rabi periods $T_R$. The~initial photonic excitation $|001\rangle$ is coherently exchanged with the symmetric electronic states $|100\rangle$ and $|010\rangle$, while $|111\rangle$ remains unpopulated. The~total probability is conserved at all times. 
\textbf{Bottom panel}: expectation values of the cavity and electronic energy contributions, showing periodic energy exchange consistent with Rabi oscillations.}
    \label{fig:landa_vs_things_0}
\end{figure}
\unskip

\subsection{Simulation of a Measured Light--Matter~System}

In this section, we present a numerical simulation mimicking experiments that
exhibit partition noise---historically invoked as evidence for the particle-like
nature of light---and show that their results can be consistently explained
without attributing any ontology to photons. The~stochasticity of the measurement process (a photon being measured sometimes at $x_1$ and sometimes at $x_2$) is provided by the stochasticity of the initial Bohmian particle positions (the quantum equilibrium described in Section~\ref{sec5}). The~anti-correlation in the detection events (a photon measured either at $x_1$ or at $x_2$, but~not at both places) arises from the non-overlapping of such a process in the enlarged configuration space constructed when we include the measuring degrees of~freedom. 

First, let us discuss what it means to detect a photon. It means to measure its energy $\hbar\omega_c$.  Ultimately, any measuring apparatus is made of matter and its pointer is linked to positions in real space. We have mentioned in the previous section that, without~the resonant condition $E_{1,p}-E_{0,p}=\hbar\omega_c=E_{1,e}-E_{0,e}=0.105$ eV, light and matter do not interact (they evolve independently), and~therefore we would be unable to measure any property of the light. Thus, the~measurement must proceed by first transferring the energy of the light to the electrons located at $x_1$ and $x_2$ under resonant conditions (as we have done in the previous non-measured simulation), and~then measuring (the increment of) the energies of the electrons at $x_1$ and $x_2$.

\subsubsection{Model   With Measuring Apparatus}

We consider a simulated quantum--optical setup depicted in Figure~\ref{fig:set_up}. This setup is the same as that in Figure~\ref{fig:set_up_non_measured}, but~with two additional measurement devices, each one measuring the energy of each one of the electrons. The~measurement devices are modeled as matter particles with coordinates $y$ and $z$ and masses $m_y$ and $m_z$ (we choose \mbox{$m_y=m_z=m_e$}). The~complete system is described by the configuration variables $(\xi,y,z)=(x_1,x_2,q,y,z)$, where $x_1$ and $x_2$ denote the electronic coordinates in each quantum well, $q$ is the coordinate of the single cavity mode, and~$y,z$ are the pointer coordinates associated with the two~detectors measuring the electronic energies. The~total Hamiltonian is therefore
\begin{equation}
    H = H_0 + H_{\mathrm{int}} + H_{\mathrm{meas}}.
    \label{Htotalmea}
\end{equation}
The free Hamiltonian (including the kinetic energy of the pointers) reads as

\begin{align}
H_0 =&
-\frac{\hbar^2}{2m_e}\frac{\partial^2}{\partial x_1^2} + V(x_1)
- \frac{\hbar^2}{2m_e}\frac{\partial^2}{\partial x_2^2} + V(x_2)
\nonumber\\
&+ \frac{\hbar\omega_c}{2}\left( -\frac{\partial^2}{\partial q^2} + q^2 \right)
- \frac{\hbar^2}{2m_y}\frac{\partial^2}{\partial y^2}
- \frac{\hbar^2}{2m_z}\frac{\partial^2}{\partial z^2}.
\end{align}
The light--matter interaction is modeled (as before) as
\begin{equation}
H_{\mathrm{int}} = \alpha q x_1 + \alpha q x_2.
\end{equation}
The measurement of the electronic energies is implemented via von Neumann-type interactions between each electron and its corresponding pointer, as~described in Section~\ref{sec5}, given~by
\begin{equation}
H_{\mathrm{meas}} =
\mu(t) P_y K_{e_1}
+
\mu(t) P_z K_{e_2},
\end{equation}
where $P_y=-i\hbar \frac{\partial}{\partial y}$ is the momentum operator of the pointer $y$ and $P_z=-i\hbar \frac{\partial}{\partial z}$ is the momentum operator of the pointer $z$. We also define $K_{e_i}=-\frac{\hbar^2}{2m_e}\frac{\partial^2}{\partial x_i^2}-E_{0,e}$ as the kinetic-energy operator of electron $x_i$ (minus its ground energy value $E_{0,e}$), which is the observable we want to measure. The~interaction of the pointer with the electrons is modeled through a time-dependent Gaussian coupling $\mu(t) = \mu_0
\exp\!\left[-{(t - t_0)^2}/{4\sigma_{\mu}^2}\right]$ with $\mu_0 = 200\,\mathrm{nm/eV/fs}$. The~Gaussian profile is centered such that the measurement devices are activated at the instant of maximum energy transfer (see Figure~\ref{fig:set_up}). This corresponds to $t_0 = \frac{\pi}{\Omega_R}$ which represents half a Rabi oscillation period. We find that a measurement width of $\sigma_{\mu} = 2\,\mathrm{fs}$ provides a satisfactory compromise between temporal resolution and smooth Bohmian dynamics\footnote{The measurement duration $\sigma_{\mu}$ must be chosen much smaller than the Rabi period in order to ensure that the measurement is performed near the desired quantum state. If~the measurement time were comparable to the Rabi period, the~interaction would probe the system throughout its dynamical evolution, leading to a less well-defined measurement outcome.
On the other hand, the~measurement cannot be made arbitrarily short. An~idealized Dirac-delta-like interaction would induce unphysical features in the Bohmian trajectories, such as divergent velocities at the instant of measurement.}.

\vspace{-4pt}
\begin{figure}[h!]
   % \centering
    \includegraphics[width=0.9\linewidth]{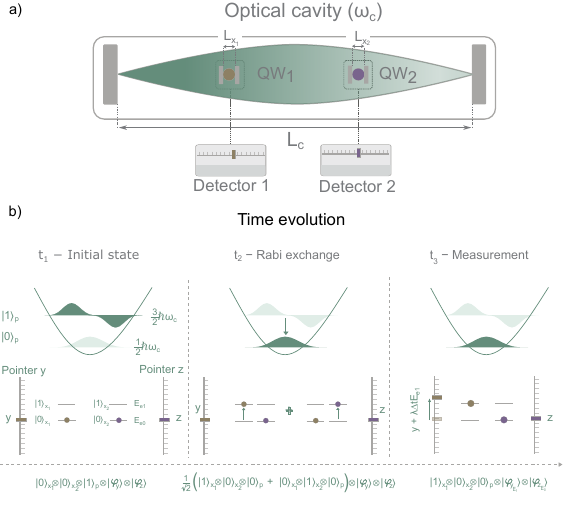}
    \caption{(\textbf{a}) Sketch
 of the simulated system including the measurement apparatus. 
An optical cavity containing a single electromagnetic mode of frequency $\omega_c$ interacts with two quantum dots (QD). 
Each electron is coupled to an independent pointer degree of freedom, represented by the coordinates $y$ and $z$, which act as measurement devices for the electronic energies.
(\textbf{b}) Two-level energy diagram illustrating the measurement process during the Rabi dynamics. 
As in the non-measured case, the~initial state at time $t_1$, $\ket{001}= \ket{0}_{x_1}\otimes\ket{0}_{x_2}\otimes\ket{1}_{p}$, evolves coherently into the superposition $\frac{1}{\sqrt{2}}\left(\ket{100} + \ket{010}\right)$ at time $t_2$. 
When the measurement interaction is activated, the~electronic states become correlated with the pointer degrees of freedom. 
This correlation dynamically separates the corresponding branches in configuration space, producing an effective collapse of the conditional wave function: depending on the Bohmian trajectory, the~system evolves either to $\ket{100}$ or to $\ket{010}$, corresponding to the energy being absorbed by electron 1 or electron 2, respectively.}
    \label{fig:set_up}
\end{figure}

The Schrödinger equation for the total wave function $\Phi\equiv\Phi(x_1,x_2,q,y,z,t)$ governs the full unitary evolution of the system and reads as

\begin{align}
\label{schomeasure}
    i\hbar \frac{\partial\Phi}{\partial t}&=\left( -\frac{\hbar^2}{2m_e}\frac{\partial^2}{\partial x_1^2} + V(x_1)\right)\Phi 
    +\left(-\frac{\hbar^2}{2m_e}\frac{\partial^2}{\partial x_2^2} + V(x_2)\right)\Phi  \\
    &+  \frac{\hbar\omega_c}{2}\left(-\frac{\partial^2}{\partial q^2} + q^2\right)\Phi+ \alpha q (x_1+x_2)\Phi \nonumber \\ 
    &+\left( -\frac{\hbar^2}{2m_y}\frac{\partial^2}{\partial y^2} 
    + i\hbar\mu(t)\frac{\hbar^2}{2 m_e} \frac{\partial}{\partial y}\frac{\partial^2}{\partial x_1^2} \right)\Phi +\left( -\frac{\hbar^2}{2m_z}\frac{\partial^2}{\partial z^2} 
    + i\hbar\mu(t)\frac{\hbar^2}{2 m_e} \frac{\partial}{\partial z}\frac{\partial^2}{\partial x_2^2} \right)\Phi.\nonumber
\end{align}
To solve the Schrödinger equation, Equation \eqref{schomeasure}, we expand the total wave function in a truncated product basis,
\begin{equation}
\Phi(x_1,x_2,q,y,z,t)= \sum_{n,m,k,=0}^{1,1,1}\sum_{l=-\mathcal{L}}^{\mathcal{L}}\sum_{s=-\mathcal{S}}^{\mathcal{S}}c_{nmkls}(t)\phi_n(x_1)\phi_m(x_2)\psi_k(q)\varphi_l(y)\varphi_s(z),
\end{equation}
where $\varphi_l$, $\varphi_s$ are the eigenfunctions of the pointer degrees of freedom, which are plane waves. The~initial coefficients $c_{nmkls}(0)$ are chosen so that their dependence
on the pointer indices, $l=-\mathcal{L},\ldots,0,\ldots,\mathcal{L}$ and $s=-\mathcal{S},\ldots,0,\ldots,\mathcal{S}$, generates Gaussian wave packets for the pointer
states $\varphi(y)$ and $\varphi(z)$. The~truncation parameters for the measurement pointers are chosen as $
\mathcal{L}=\mathcal{S}=10$ (see Appendix~\ref{ap:meas} for more
details). 

Substituting this expansion into \eqref{schomeasure} and projecting onto
the basis states yields coupled equations of motion for the
coefficients $c_{nmkls}(t)$ completely equivalent to that of the {non-measured case \eqref{eq:c_nmk}, but~with a modified diagonal energies given by the sum  $E_{nmkls}
=
E_{n,e}
+
E_{m,e}
+
E_{k,p}
+
E_{l,y}
+
E_{s,z}$}. In~any case, the~structure of the interaction terms is not modified by the addition of the measurement devices, so the Hamiltonian still preserves the total parity mentioned in \eqref{up}. The~two Hilbert sub-spaces are
\begin{equation}
\mathcal{H}_{even}
=
\mathrm{span}
\{ {\ket{000},\ket{110},\ket{101},\ket{011}} \}\otimes
\mathcal H_{(l,s)},
\end{equation}
and
\begin{equation}
    \mathcal{H}_{odd}
=
\mathrm{span}
\{\ket{100},\ket{010},\ket{001},\ket{111} \}\otimes
\mathcal H_{(l,s)},
\end{equation}
with
\begin{equation}
\mathcal H_{(l,s)}
=
\mathrm{span}
\left\{
\ket{l,s}
\right\}.
\end{equation}
As in the previous section, the~unitary dynamics guarantee that the state
remains in $\mathcal{H}_{odd}$ for all times, and~no amplitude is ever
transferred to $\mathcal{H}_{even}$. For~this reason, numerical simulations
can be restricted to $\mathcal{H}_{odd}$ without loss of~generality.

\subsubsection{Numerical~Results  With Measuring Apparatus}

The results presented here are obtained from the wave function $\Phi(x_1,x_2,q,y,z,t)$, which is the solution of the Schrödinger equation, Equation \eqref{schomeasure}. In~addition, different Bohmian trajectories defined as
\begin{equation}
\zeta^{(\eta)}(t)\equiv (\xi^{(\eta)}(t),y^{(\eta)}(t),z^{(\eta)}(t))\equiv(x^{(\eta)}_1(t),x^{(\eta)}_2(t),q^{(\eta)}(t),y^{(\eta)}(t),z^{(\eta)}(t)),
\end{equation}
can also be computed, as~explained in the Bohmianization in Section~\ref{sec4} (see Appendix~\ref{ap:trajectories} for concrete details of this simulation). Notice that $\xi^{(\eta)}(t)$ includes the matter and light degrees of freedom, but~not the pointers' degrees of freedom, while $\zeta^{(\eta)}(t)$ includes all five degrees of freedom. The~superscript $\eta$ labels different experiments, and~the trajectories in different experiments are selected according to the quantum equilibrium hypothesis, as~discussed in Section~\ref{sec5}.

In particular, we consider $\xi^{(1)}(t)$ corresponding to a detection of a photon at $x_1$ (strictly speaking, to~a trajectory where the pointer $y^{(1)}(t)$ moves appreciably far from its initial central position), and~$\xi^{(2)}(t)$ corresponding to a detection of a photon at $x_2$ (strictly speaking, to~a trajectory where the pointer $z^{(2)}(t)$ moves appreciably far from its initial central position).

We recall that both Bohmian trajectories evolve under the same total wave function $\Phi(x_1,x_2,q,y,z,t)$. However, this wave function can be conditioned on different Bohmian trajectories, leading to different conditional wave functions. For~example, using the previous two trajectories, we define the conditional wave function discussed in \eqref{wavemeas} as
\begin{equation}
\Psi^{(\eta)}(\xi,t)\equiv \Psi^{(\eta)}(x_1,x_2,q,t)= \Phi(x_1,x_2,q,y^{(\eta)}(t),z^{(\eta)}(t),t) \quad \text{for} \quad \eta=1,2.
\label{funcon}
\end{equation}
Here, $\Psi^{(1)}(x_1,x_2,q,t)$ corresponds to a detection of a photon at $x_1$, whereas $\Psi^{(2)}(x_1,x_2,q,t)$ corresponds to a detection of a photon at $x_2$.

Figure~\ref{fig:things_vs_landa} shows the evolution of the (conditional) probabilities $|c^{(\eta)}_{nmk}|^2$, for~$\eta=1,2$, computed by projecting the conditional wave function $\Psi^{(\eta)}(x_1,x_2,q,t)$, defined in \eqref{funcon}, onto~the product basis $\phi_n(x_1)\phi_m(x_2)\psi_k(q)$ as
\begin{equation}
c^{(\eta)}_{nmk}(t)=\int dx_1\,dx_2\,dq\;\phi_n^*(x_1)\phi_m^*(x_2)\psi_k^*(q)\Psi^{(\eta)}(x_1,x_2,q,t).
\end{equation}
It also plots the (conditional) expectation values of the different terms $H_{j}$ of the Hamiltonian in \eqref{Htotalmea}, defined as
\begin{equation}
\langle H^{(\eta)}_{j}(t) \rangle =\int dx_1\,dx_2\,dq\; \Psi^{(\eta)*}(x_1,x_2,q,t)\, H_{j}\, \Psi^{(\eta)}(x_1,x_2,q,t).
\end{equation}

The measurement in Figure~\ref{fig:things_vs_landa} is activated around $t_3\approx57\,\mathrm{fs}$. Before~the measurement takes place, the~degrees of freedom $x_1,x_2,q$ evolve independently of $y$ and $z$, and~both the coefficients and the energies evolve in the same way as in the case without measuring apparatus shown in Figure~\ref{fig:landa_vs_things_0}, displaying the coherent Rabi oscillations of a closed~system.

\begin{figure}[h!]
    \centering    \includegraphics[width=1\linewidth]{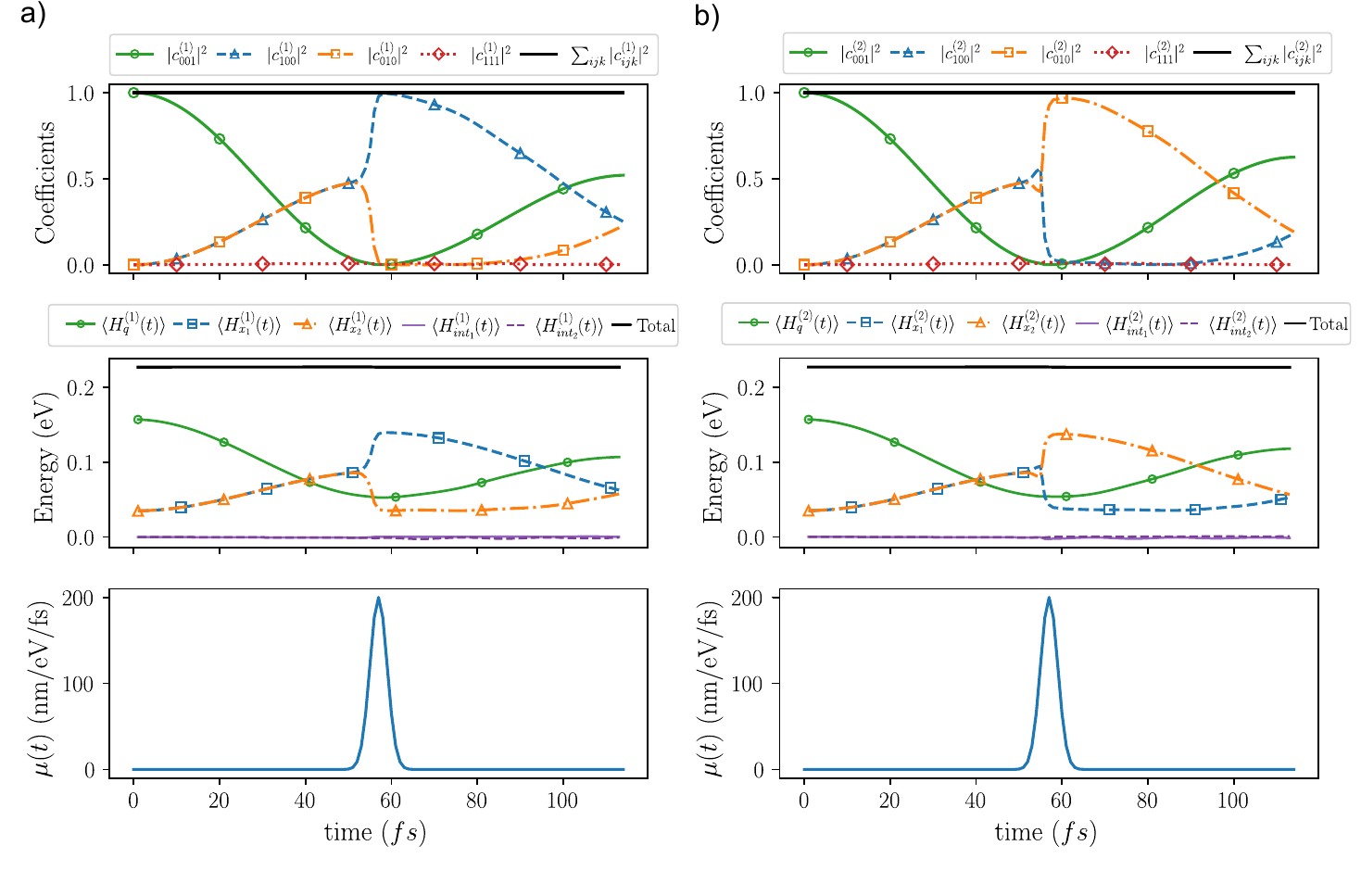}
    \caption{Measurement
-induced effective collapse in the two possible Bohmian outcomes. 
\linebreak  (\textbf{a})
 Evolution of the conditional state and corresponding expectation values along a trajectory that follows the $y$-pointer branch, leading to $|c_{100}|^2 \to 1$ and $\langle H_{x_1} \rangle \to E_{1,e}$, corresponding to excitation of the first electron
 . 
(\textbf{b}) Evolution along the alternative trajectory that follows the $z$-pointer branch, yielding $|c_{010}|^2 \to 1$ and $\langle H_{x_2} \rangle \to E_{1,e}$, corresponding to excitation of the second electron%Please check intended meaning has been retained
. 
In both cases, the~total evolution remains unitary, while the effective collapse emerges from the dynamical separation of the pointer wave packets.
}
    \label{fig:things_vs_landa}
\end{figure}
%\unskip

\pagebreak

During the measurement, the~dynamics change dramatically. In~the left panels of \mbox{Figure~\ref{fig:things_vs_landa}}, the~($\xi^{(1)}$-conditional) probability $|c^{(1)}_{100}(t)|^2$ approaches unity and the corresponding energy $\langle H^{(1)}_{x_1}(t)\rangle$ reaches the value $E_{1,e}=0.140$ eV. This corresponds to the trajectory $\xi^{(1)}(t)$, whose pointer coordinate moves along the $y$ direction. Just after the measurement, the~($\xi^{(1)}$-conditional) state is clearly defined by the evolution in \eqref{sup1}, giving the measured quantum state $\ket{100}$. In~the right panels of Figure~\ref{fig:things_vs_landa}, the~($\xi^{(2)}$-conditional) probability $|c^{(2)}_{010}(t)|^2$ approaches unity and the corresponding energy $\langle H^{(2)}_{x_2}(t)\rangle$ reaches the value $E_{1,e}=0.140$ eV. This corresponds to the trajectory $\xi^{(2)}(t)$, whose pointer coordinate moves along the $z$ direction. Just after the measurement, the~($\xi^{(2)}$-conditional) quantum state is clearly defined by the evolution in \eqref{sup2}, giving the measured\footnote{The measurement itself can be regarded as a preparation procedure that sets new initial conditions for the system, different from the state $\ket{001}$ used in the non-measured simulations. After~the measurement (i.e., preparation), the~coherent evolution of the light--matter system with Rabi oscillations occurs again.}  quantum state $\ket{010}$.

To help understand the full measurement process, we plot several moduli of the conditional wave functions in Figure~\ref{fig:cond_wf}. In~the top-left panel we plot the modulus square of the conditional wave function $\Psi^{(1)}(y,z,t)$, whose initial shape is $\Psi^{(1)}(y,z,0) \propto \varphi(y)\varphi(z)$ and which during the measurement at time $t_3$ becomes $\Psi^{(1)}(y,z,t_3)=\varphi(y-y_{E_1})\varphi(z-z_{E_0})$ with $y_{E_1}=(E_{1,e}-E_{0,e})\int_{t_0}^{T_{\mathrm{sim}}}\mu(t)\,dt$ and $z_{E_0}=(E_{0,e}-E_{0,e})\int_{t_0}^{T_{\mathrm{sim}}}\mu(t)\,dt=0$. Similarly, in~the top-right panel of Figure~\ref{fig:cond_wf} we plot the modulus square of the conditional wave function $\Psi^{(2)}(y,z,t)$ that  during the measurement at time $t_3$ becomes $\Psi^{(2)}(y,z,t_3)\propto\varphi(y-y_{E_0})\varphi(z-z_{E_1})$, with~$z_{E_1}=(E_{1,e}-E_{0,e})\int_{t_0}^{T_{\mathrm{sim}}}\mu(t)\,dt$ and   $y_{E_0}=(E_{0,e}-E_{0,e})\int_{t_0}^{T_{\mathrm{sim}}}\mu(t)\,dt=0$. In~summary, the~(conditional) measured wave functions $\Psi^{(1)}(y,z,t)$ and $\Psi^{(2)}(y,z,t)$ have different movements indicating the output results: the first pointer trajectory moves horizontally ($y$ axis), while the second pointer trajectory moves vertically ($z$ axis).

In Figure~\ref{fig:cond_wf}, we also plot the modulus of $\Psi^{(1)}(x_1,x_2,t)$, which corresponds to $\ket{00}$ at the initial time, evolves to $\frac{1}{\sqrt{2}}(\ket{10}+\ket{01})$ at time $t=t_2$, and~becomes $\ket{10}^{(1)}$ at time $t=t_3$. Abusing a bit of the language, the~notation $\ket{10}^{(1)}$ wants to refer to $\Psi^{(1)}(x_1,x_2,t)=\phi_1(x_1)\phi_0(x_2) \psi_k(q^{(1)}(t))\varphi_l(y^{(1)}(t))\varphi_s(z^{(1)}(t))$. The~same evolution happens for $\Psi^{(2)}(x_1,x_2,t)$, but~the final result is $\ket{01}^{(2)}$ when measured. The~additional plots of the modulus of $\Psi^{(\eta)}(x_1,q,t)$ and $\Psi^{(\eta)}(x_2,q,t)$ corroborate the evolutions just described. Therefore, we conclude that when the pointer $y^{(1)}(t)$ moves while $z^{(1)}(t)$ does not, the~conditional wave function $\Psi^{(1)}(x_1,x_2,q,t)$ corresponds to $\ket{100}^{(1)}$. Conversely, when the pointer $z^{(2)}(t)$ moves while $y^{(2)}(t)$ does not, the~conditional wave function $\Psi^{(2)}(x_1,x_2,q,t)$ corresponds to $\ket{010}^{(2)}$.

The key point in understanding the process of measurement is that $\ket{100}$ and $\ket{010}$ are, initially, in a coherent superposition (i.e., overlap) in the configuration space $(x_1,x_2,q)$ of the non-measured simulations, but~such superposition is \textit{naturally} broken when measured because the (conditional) states $\ket{100}^{(1)}$ and $\ket{010}^{(2)}$ do not longer overlap because they \textit{live} in an enlarged configuration space $(x_1,x_2,q,y,z)$. As~discussed in Section~\ref{sec5}, $\ket{100}^{(1)}$ and $\ket{010}^{(2)}$ belong to different (non-overlapping) branches and the Bohmian trajectory, with~its well-defined coordinates, can only be in one of the branches\footnote{Of course, if~one prefers to ignore the detailed discussion of the roles played by $y^{(\eta)}(t)$ and $z^{(\eta)}(t)$ in breaking the superposition, one may still explain the detection of the photon by invoking a (non-unitary and stochastic) fundamental collapse process such as $\frac{1}{\sqrt{2}}(\ket{100}+\ket{010}) \to \ket{100}$ or $\frac{1}{\sqrt{2}}(\ket{100}+\ket{010}) \to \ket{010}$, depending on the outcome of ``God's dice''. Although~the mathematical description of such a collapse is certainly simpler, it appears far less satisfactory from a fundamental point of view.}.

\begin{figure}[h!]
    \centering
    \includegraphics[width=1\linewidth]{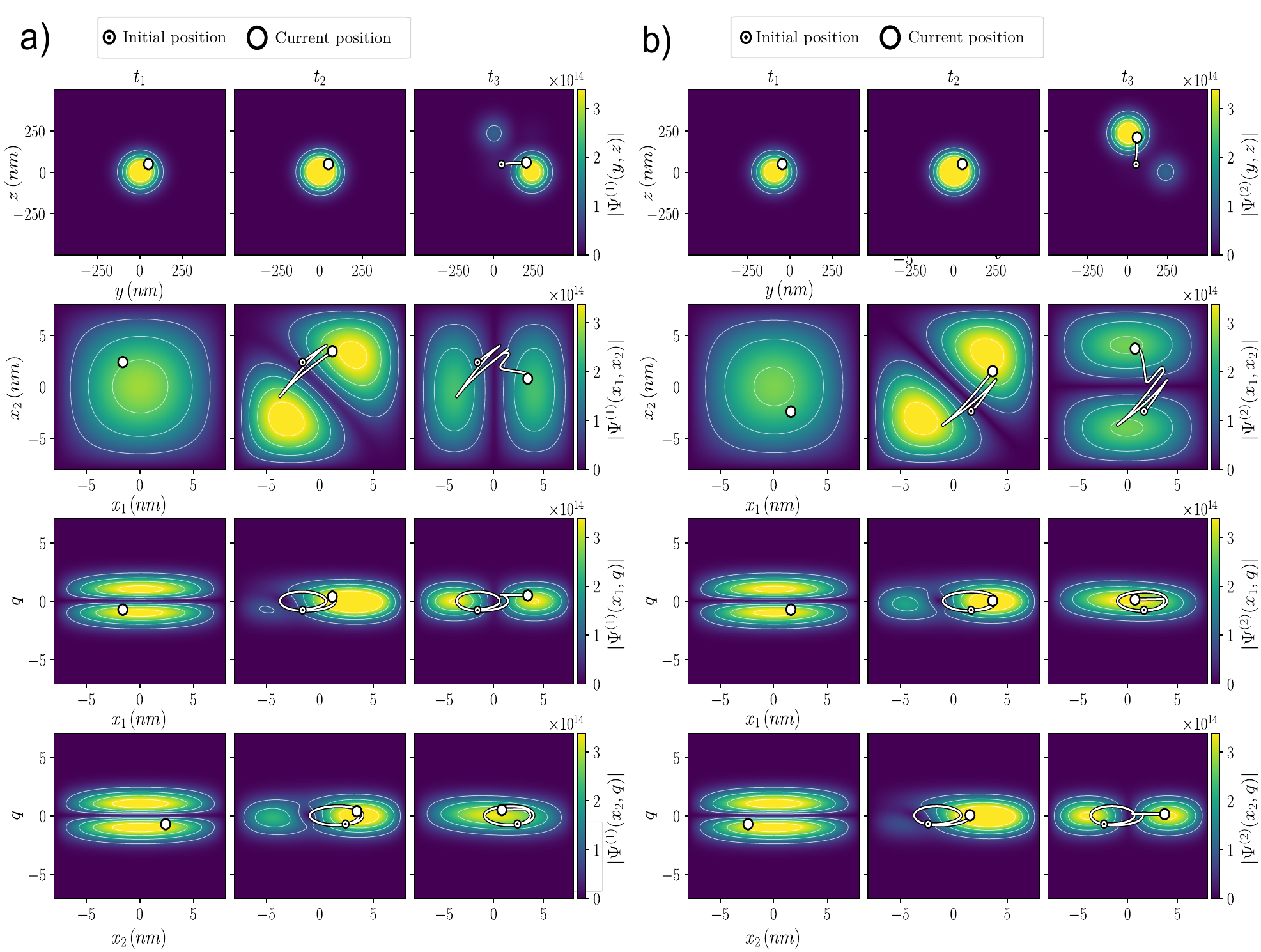}
\caption{Conditional
 wave functions and corresponding Bohmian trajectories at three different times. 
At $t_1$, the~system is in the initial state of the simulation,
$
\ket{\Phi(t_1)}=\ket{001}\otimes \ket{\varphi_y} \otimes \ket{\varphi_z}.
$
At $t_2$, corresponding to half a Rabi oscillation, the~energy transfer is maximal and the state has evolved into the superposition
$
\ket{\Phi(t_2)}=\frac{1}{\sqrt{2}}\left(\ket{100} + \ket{010}\right)\otimes \ket{\varphi_y} \otimes \ket{\varphi_z}.
$
At the later time $t_3$, the~measurement interaction has been completed. 
(\textbf{a})
 The~Bohmian trajectory follows the wave packet associated with pointer $y$, indicating that the energy has been absorbed by the first electron
 . 
(\textbf{b}) The~trajectory follows the wave packet associated with pointer $z$, corresponding to the case in which the energy is absorbed by the second electron
}
    \label{fig:cond_wf}
\end{figure}

We have therefore shown how partition noise (i.e., the~photon is detected at $x_1$ or at $x_2$, but~never at both positions) arises from the measurement of the electronic energies, without~requiring photons to be interpreted as ontic discrete corpuscular~entities.

In the non-measured simulation, the~energy is transferred symmetrically to both electrons as a consequence of the coherent unitary dynamics. Only when measurement devices are included does an effective collapse emerge through the dynamical branching of the wave function, leading to definite outcomes linked to the Bohmian trajectories of the pointers. Under~the condition that the available energy is sufficient to excite only one electron at a time, partition noise naturally appears in the statistics of measurement outcomes\footnote{Throughout the simulation, the~dynamics are restricted to the sub-space $\mathcal{H}_{\mathrm{odd}}$, which contains states with an odd number of excitations. The~structure of the Hamiltonian therefore prevents transitions to states such as $\ket{110}$, $\ket{101}$, or~$\ket{011}$. In~principle, the~state $\ket{111}$ belongs to the same parity sector and is dynamically allowed. However, transitions such as $\ket{100} \to \ket{111}$ require deep strong coupling, which occurs only when the interaction term $\alpha x q$ becomes comparable in magnitude to the free Hamiltonian contributions. In~that deep strong regime (not considered in this paper), energy stored in the interaction can be redistributed to the rest of the system. Importantly, this behavior remains fully consistent with energy conservation and arises purely from the unitary dynamics of the model.}.

\section{Conclusions}
\label{conc}

By revisiting and rehabilitating earlier Bohmian proposals~\cite{Bohm1952II,Bohm1987,Belinfante1973,Holland,Kaloyerou1994}, we show that cavity quantum electrodynamics experiments can be modeled using Bohmian electron trajectories in physical space together with well-defined electromagnetic fields, represented by appropriate evolving variables. We explicitly discuss how the Born rule is satisfied, and~we numerically simulate an experiment involving photon partition~noise.

The paper presents two main innovations. The first is pedagogical, with~implications for numerical computations. The~claim that deterministic trajectories governed by a unitary Schrödinger equation cannot account for processes involving photon creation and annihilation is largely semantic and historically conditioned. We show that processes described in Fock space as photon creation and annihilation can be more simply reinterpreted as quantized energy exchanges between matter and electromagnetic field modes\footnote{Even the creation and annihilation of electrons can be interpreted as a semantic description of a change in the electron’s energy from a negative value (in the Dirac sea) to a positive one, or~vice~versa, within~a “persistent particle ontology”  as described in~\cite{Deckert2019} or {when the ``Dirac sea is taken seriously'' as proposed in~\cite{ColinStruyve}}.}. From~this perspective, unitary Schrödinger evolution in coordinate space provides a conceptually straightforward framework to discuss practical aspects of quantum electrodynamics, relying on the same mathematical machinery used in standard non-relativistic quantum~mechanics. 

The use of Bohmian trajectories to solve the dynamics of such systems enables a discussion of quantum electrodynamics within a well-defined world in which particles and fields evolve non locally, continuously, and~deterministically, reproducing the statistical predictions of quantum theory while offering a transparent account of individual processes. Beyond~providing a clear ontology, this framework explains how quantum electrodynamics measurements can be understood —specifically, how the superposition principle is effectively broken during measurement—without invoking any collapse postulate. The~overall message is that a physical theory with a well-defined ontology helps to better understand physical phenomena, free from ambiguities and apparent paradoxes. This paper also serves as a presentation of a software package, QC-Slim~\cite{QCslim}, that has been developed following these pedagogical principles, and~it opens new avenues for the numerical computation of quantum electrodynamic~phenomena. 

The second major contribution of this paper is a detailed analysis of what is actually measured when a photon is detected. We argue that, in~photon-detection events, what is experimentally recorded is the motion of matter in the apparatus. We show that photon partition noise does not require photon trajectories, i.e., photon partition noise is not a manifestation of an intrinsic particle nature of light, but~rather a consequence of the particle nature of the material~pointers.

This discussion is particularly relevant for understanding experiments that measure momentum and position for photons to evaluate empirically a weak value~\cite{Kocsis2011,Oriols2025}. Such weak values have been interpreted as empirical determinations of Bohmian photon velocities. However, within~the traditional Bohmian mechanics presented in this paper there is no such notion as a photon velocity in the ontological sense. Therefore, the~present paper offers also a path to clarify what is actually being measured in such type of~experiments.

In this paper, we also suggest the possibility (already indicated in the literature~\cite{Schwarzschild1903,Tetrode1922,Fokker1929,Feynman1945,Feynman1949,Dustin2018, Deckert_thesis}) of interpreting quantum electrodynamical phenomena within a Bohmian trajectory fermion-only ontology, in~which both the photon concept and the electromagnetic field itself are regarded as emergent rather than fundamental (ontic) entities, due to other non-simulated electrons. Within~this Bohmian trajectory fermion-only ontology, the~translation of the classical results in \eqref{chargeden} and \eqref{currentden} is straightforward, requiring only the replacement of classical electron trajectories with Bohmian ones. The~expressions in \eqref{retardedpotentialsca} and \eqref{retardedpotentialvec} can then be directly applied. The~results presented in Section~\ref{sec6}, where photon properties are measured while being defined as non-ontic quantities from the beginning, show that only ontic trajectories for the matter pointers are required for the traditional Bohmian explanation of quantum measurements of individual~systems.

%%%%%%%%%%%%%%%%%%%%%%%%%%%%%%%%%%%%%%%%%%

%%%%%%%%%%%%%%%%%%%%%%%%%%%%%%%%%%%%%%%%%%
\vspace{6pt} 

%%%%%%%%%%%%%%%%%%%%%%%%%%%%%%%%%%%%%%%%%%
%% optional
%\supplementary{The following supporting information can be downloaded at:  \linksupplementary{s1}, Figure S1: title; Table S1: title; Video S1: title.}


% Only used for preprtints:
% \supplementary{The following supporting information can be downloaded at the website of this paper posted on \href{https://www.preprints.org/}{Preprints.org}.}

%\begin{tabularx}{\textwidth}{lll}
%\toprule
%\textbf{Name} & \textbf{Type} & \textbf{Description} \\
%\midrule
%S1 & Python script (.py) & Script of python source code used in XX \\
%S2 & Text (.txt) & Script of modelling code used to make Figure X \\
%S3 & Text (.txt) & Raw data from experiment X \\
%S4 & Video (.mp4) & Video demonstrating the hardware in use \\
%\bottomrule
%\end{tabularx}
%}

%%%%%%%%%%%%%%%%%%%%%%%%%%%%%%%%%%%%%%%%%%

\paragraph{Funding}This research was funded by Spain’s
Ministerio de Ciencia, Innovación y Universidades under
Grants PID2021-127840NB-I00 (MICINN/AEI/FEDER,
UE), PID2024-161603NB-I00,  and~PDC2023-145807-I00 (MICINN/AEI/FEDER,
UE), and~European Union’s
innovation program under grant ID:101084035~DigiQ. 

\paragraph{Acknowledgments}This work has greatly benefited from discussions with Xabier
Oianguren-Asua. Numerical experiments and simulations were implemented in
Python~3.12.

\appendix

\renewcommand{\theequation}{\thesection.\arabic{equation}}
\setcounter{equation}{0}

%\section{Creation and Annihilation in Fock~Space}
\section[\appendixname~\thesection. Creation and Annihilation in Fock~Space]{Creation and Annihilation in Fock~Space}
\label{appendix}

Instead of the operators  $\hat \iqq_\lambda = \sqrt{\hbar} \iqqq_\lambda$ and $\hat \ipp_\lambda = -i\sqrt{\hbar }\frac{\partial}{\partial \iqqq_\lambda}$ in \eqref{change2} to define $ H_{\text{field},\lambda}$ in \eqref{hfield2}, one can invoke different operators, named annihilation and creation operators, respectively,
\begin{equation}
\hat{a}_\lambda = \frac{1}{\sqrt{2}}\left(\frac{\partial}{\partial \iqqq_\lambda}+\iqqq_\lambda\right),
\qquad
\hat{a}^\dagger_\lambda = \frac{1}{\sqrt{2}}\left(-\frac{\partial}{\partial \iqqq_\lambda}+\iqqq_\lambda\right).
\end{equation}
that satisfy $\hat{a}^\dagger_\lambda \hat{a}_\lambda=\frac{1}{2}\left( -  \frac{\partial^2}{\partial \iqqq_\lambda^2}+ \iqqq_\lambda^2-1 \right)$ and $\hat{a}_\lambda \hat{a}^\dagger_\lambda=\frac{1}{2}\left( -  \frac{\partial^2}{\partial \iqqq_\lambda^2}+ \iqqq_\lambda^2+1 \right)$ so that $[\hat{a}_\lambda,\hat{a}^\dagger_\lambda]=1$. With~these new operators, the~energy of one mode of the electromagnetic field can be rewritten as
\begin{equation}
H_{\text{field},\lambda}=\frac{\hbar \omega_\lambda}{2}\left(-  \frac{\partial^2}{\partial \iqqq_\lambda^2}+ \iqqq_\lambda^2\right)
=
\hbar\omega_\lambda
\left(
\hat{a}^\dagger_\lambda \hat{a}_\lambda
+
\frac{1}{2}
\right).
\end{equation}
Then, one can straightforwardly show that $[\hat{a}_\lambda,H_{\text{field},\lambda}]=\hbar \omega_\lambda \hat{a}_\lambda$ and  $[\hat{a}^\dagger_\lambda,H_{\text{field},\lambda}]=-\hbar \omega_\lambda \hat{a}^\dagger_\lambda$. Finally, the~energy eigenstates of the $H_{\text{field},\lambda}$ in \eqref{light_eigenstates} can be written here as the ``ket'' corresponding to $m$ photons by the ``bra'' of the coordinate $\iqqq_\lambda$:
\begin{equation}
\psi_{m,\lambda}(\iqqq_\lambda)\equiv \langle \iqqq_\lambda| m\rangle.
\end{equation} 
Then, one can identify the annihilation operator
\begin{equation}
\hat a_\lambda \, |m\rangle = \sqrt{m} \, |m-1\rangle,
\end{equation}
as the operator the eliminates one photon, while the creation operator
\begin{equation}
\hat a^\dagger_\lambda \, |m\rangle = \sqrt{m+1} \, |m+1\rangle.
\end{equation}
is the one creating a photon. In~particular,
\begin{equation}
\hat a_\lambda \, |0\rangle = 0,
\end{equation}
since the vacuum contains no particles. If~we assimilate the photon as a particle then we are dealing with the states $|0\rangle, |1\rangle,|2\rangle,...$ with a different number of particles. Such an interpretation finds a natural definition in the so-called Fock Space $\mathcal{F}$ to deal with a variable number of identical~particles.  

In this new language, we can re-interpret the wave function $\Psi(\mathbf{r},\iqqq_{1},\ldots,\iqqq_{\mathcal M},t)$ in \eqref{wavefunction} as a superposition of different electromagnetic fields with a different number of photons in each electromagnetic mode, and~the modulus square of $c_{n,m_1,...,m_{\mathcal{M}}}(t)$ in \eqref{component} as a the probability of having $m_1$ photons in the first mode, $m_2$ in the second and so. The~evolution of such superposition with a different number of particles will be exactly given by \eqref{schoc} with $\beta_{n,m_1,...,m_{\mathcal{M}};n',m'_1,...,m'_{\mathcal{M}}}$ and $\gamma_{n,m_1,...,m_{\mathcal{M}};n',m'_1,...,m'_{\mathcal{M}}}$ providing the transition between configurations with a different number of~particles.

%\section{Schrödinger Equation for Two Electrons and One Electromagnetic~Mode}
\section[\appendixname~\thesection. Schrödinger Equation for Two Electrons and One Electromagnetic~Mode]{Schrödinger Equation for Two Electrons and One Electromagnetic~Mode}
\label{ap:no_meas}

The simplifications mentioned in the introduction of Section~\ref{sec611} straightforwardly leads to the following Schrödinger equation:
\begin{align}
i\hbar \frac{\partial \Psi}{\partial t}
=&
\left(
-\frac{\hbar^2}{2m_e}\frac{\partial^2}{\partial x_1^2}
+ V(x_1)
\right)\Psi
+
\left(
-\frac{\hbar^2}{2m_e}\frac{\partial^2}{\partial x_2^2}
+ V(x_2)
\right)\Psi
\nonumber\\
&+
\frac{\hbar \omega_c}{2}
\left(
- \frac{\partial^2}{\partial q^2}
+ q^2
\right)\Psi
-
\frac{1 }{2m_e}\alpha'^2 \frac{\partial^2}{\partial \iqqq^2}\,\Psi 
+
\frac{\hbar}{m_e}\alpha' \frac{\partial}{\partial \iqqq}
\frac{\partial \Psi}{\partial x_1}
+
\frac{\hbar}{m_e}\alpha' \frac{\partial}{\partial \iqqq} \frac{\partial \Psi}{\partial x_2},
\label{eq:scho_derivative}
\end{align}
where we define
\begin{equation}
\alpha'= \frac{\qele\sqrt{\hbar} U_0}{\sqrt{\omega_c\varepsilon_0}}.
\end{equation}
To simplfy the notation, we also define  $p_j=(-i)\hbar \frac{\partial}{\partial x_j}$ with $j=1,2$ and $s=(-i)\frac{\partial}{\partial q}$. We can then write the Hamiltonian in \eqref{eq:scho_derivative} as
\begin{align}
    H= \frac{p_1^2}{2m_e} + V(x_1)+ \frac{p_2^2}{2m_e} + V(x_2)+\frac{\hbar\omega_c}{2}\left( q^2 + s^2\right) + \frac{1}{2m_e}\alpha'^2s^2 - \frac{\alpha'}{m_e}sp_1 - \frac{\alpha'}{m_e}sp_2,
    \label{eq:scho_qs}
\end{align}
If we now define the unitary operator as
\begin{equation}
    \mathbb U=\exp\left(\frac{i}{\hbar}\alpha's(x_1 +x_2)\right),
\end{equation}
we can perform a gauge change given by
\begin{equation}
    H'=\mathbb UH\mathbb U^{\dagger}+i\hbar \left(\partial_t\mathbb U\right)\mathbb U^{\dagger}, \quad \Psi'=\mathbb U\Psi.
\end{equation}
Since  $\mathbb U\neq \mathbb U(t)$, the~temporal derivative vanishes and we only need to do the transformation $\mathbb UH\mathbb U^{\dagger}$. Since the position operators $x_1,\; x_2$ are the generators of translations in the momentum $p$ space, and~$s$ is generator of translation in $\iqqq$ space, the~transformation can be easily performed:
\begin{align}
    & \mathbb U p_j\mathbb U^{\dagger}= p_j + \alpha' s
    \nonumber\\
    &\mathbb U q \mathbb U^{\dagger}= q +\frac{\alpha'}{\hbar}
    (x_1+x_2) ,
\end{align}
 in  \eqref{eq:scho_qs}. The~transformation on the electrons positions operators leave them unchanged, $\mathbb Ux \mathbb U^{\dagger}= x$, and the same is true for $s$. 
So, applying the transformation gets
\begin{align}
    H'= &\frac{\left(p_1 + \alpha's\right)^2}{2m_e} + V(x_1)+ \frac{\left(p_2 + \alpha's\right)^2}{2m_e} + V(x_2)+\frac{\hbar\omega_c}{2}s^2 +
    \nonumber\\
    &+\frac{\hbar \omega_c}{2}\left(q+\frac{\alpha'}{\hbar}(x_1 + x_2)\right)^2 + \frac{1}
    {2m_e}\alpha'^2s^2 - \frac{\alpha'}{m_e}s\left(p_1 + \alpha's\right) - \frac{\alpha'}{m_e}s\left(p_2 + \alpha's\right),
\end{align}
Expanding squares gets
\begin{align}
    H'&= \frac{p_1^2}{2m_e} + V(x_1) + \frac{p_2^2}{2m_e} +V(x_2) + \frac{\hbar\omega_c}{2}(q^2 + s^2)
    \nonumber\\
    &+ \alpha q (x_1+x_2) + \frac{3\alpha^2}{2m_e\omega_c^2}s^2 + \frac{\alpha^2}{2\hbar\omega_c}(x_1 + x_2)^2
\end{align}
with $\alpha=\omega_c\alpha'$. 
Note that the terms proportional to $\alpha^2 s^2$ and $\alpha^2 (x_1 + x_2)^2$ introduce corrections to the eigenenergies of the full Hamiltonian and, more generally, modify its eigenstates. Consequently, the~original product basis does not exactly diagonalize the Hamiltonian, and~these contributions generate both diagonal energy shifts and off-diagonal couplings in that~representation.

The diagonal contributions lead to small energy corrections. For~$\alpha=10^{-12}\,\mathrm{J/m}$ and $\omega_c\sim10^{14}\,\mathrm{Hz}$, which are the values used in the simulations, one obtains \mbox{$s^2\alpha^2/(\omega_c^2 m_e)\sim 10^{-21}\,\mathrm{J}$} and $\alpha^2(x_1+x_2)^2/(2\hbar\omega_c)\sim 10^{-23}\,\mathrm{J}$, which are negligible compared to the characteristic energy scale $\hbar\omega_c\sim10^{-20}\,\mathrm{J}$.

The off-diagonal contributions associated with $s^2$ induce  transitions $|k\rangle \leftrightarrow |k\pm2\rangle$ in the electromagnetic mode. Within~the two-level truncation adopted for both electrons and the field, these transitions are projected out and therefore do not contribute to the effective dynamics. The~same is true for $x_1^2$ and $x_2^2$.
In contrast, the~term proportional to $x_1 x_2$ survives the projection onto the two-level electronic sub-space. Formally, the \linebreak  $\sim$$x_1x_2$ term introduces a 'Rabi' oscillation between both electrons. The~frequency of this interaction is $\Omega_{xx}\sim \frac{\alpha^2}{\hbar\omega_c}\langle 0|x_1|1\rangle\langle 0|x_2|1\rangle=\frac{\alpha^2}{\hbar\omega_c}\langle 0|x_1|1\rangle^2$, whereas the usual Rabi frequency is $\Omega_R\sim \alpha\langle 0|x_1|1\rangle\langle 0|q|1\rangle$. Given that, under~the parameters considered in the simulations in Section~\ref{sec6}, $\langle 0|x_1|1\rangle\sim 10^{-9}$ and $\langle 0|q|1\rangle=0.707$, the~relative value between the timescales $\tau_R$ and $\tau_{xx}$ is
\begin{equation}
    \frac{\tau_{xx}}{\tau_R}=\frac{\Omega_R}{\Omega_{xx}}=\frac{\hbar\omega_c \langle 0|q|1\rangle}{\alpha  \langle 0|x|1\rangle}\sim 8,
\end{equation}
which means that the transition produced by $x_1x_2$ is eight times slower than the $x_1q$ and $x_2q$ terms. Within~the time window considered in the simulations in Section~\ref{sec6}, this term can be neglected.
 This leaves the effective Hamiltonian:
\begin{align}
    H' &= \frac{p_1^2}{2m_e} + V(x_1) + \frac{p_2^2}{2m_e} + V(x_2)
    + \frac{\hbar\omega_c}{2}(q^2 + s^2)
    + \alpha q (x_1+x_2).
\end{align}

%\section{Schrödinger Equation for Two Electrons, One Electromagnetic Mode and Two~Pointers}
\section[\appendixname~\thesection. Schrödinger Equation for Two Electrons, One Electromagnetic Mode and Two~Pointers]{Schrödinger Equation for Two Electrons, One Electromagnetic Mode and Two~Pointers}

\label{ap:meas}

The initial state corresponds to the configuration depicted in Figure~\ref{fig:set_up} which corresponds to the excited electromagnetic field state and the ground electronic states:
\begin{equation}
\ket{\Phi(0)} =
\ket{001}
\otimes
\ket{\varphi_y}
\otimes
\ket{\varphi_z}.
\end{equation}
The pointer states that $\ket{\varphi_y}=\sum_{l=-\mathcal{L}}^{\mathcal{L}} c_l \ket{l}$ and $\ket{\varphi_z}=\sum_{s=-\mathcal{S}}^{\mathcal{S}} d_s \ket{s}$ are constructed as a Gaussian superposition in the truncated momentum-like basis $\{\ket{l}\}$ and $\{\ket{s}\}$ with $\mathcal{L}=\mathcal{S}=10$. The~corresponding spatial wave functions are
\begin{equation}
\varphi(y)
=
\langle y|\varphi_y\rangle
=
\sum_{l=-\mathcal{L}}^{\mathcal{L}}
c_l \varphi_l(y),
\qquad
\varphi(z)
=
\langle z|\varphi_z\rangle
=
\sum_{s=-\mathcal{S}}^{\mathcal{S}}
d_s \varphi_s(z).
\end{equation}
with coefficients
\begin{equation}
c_l =
\mathcal{N}_y
\exp\!\left[
-\frac{(l-l_0)^2}{4\sigma_{k_y}^2}
\right],
\qquad
d_s =
\mathcal{N}_z
\exp\!\left[
-\frac{(s-s_0)^2}{4\sigma_{k_z}^2}
\right].
\end{equation}
where $\sigma_{k_y}$ and $\sigma_{k_z}$ control the width of the Gaussian, and~the discrete
Fourier expansion is
\begin{equation}
\varphi_l(y)=\langle y|l\rangle=
\frac{1}{\sqrt{L_y}}
\exp\!\left(i\pi l \frac{y}{L_y}\right),
\qquad
\varphi_s(z)=\langle z|s\rangle=
\frac{1}{\sqrt{L_z}}
\exp\!\left(i\pi s \frac{z}{L_z}\right).
\end{equation}
Thus, the~Gaussian structure in coefficient space determines the initial
shape of the pointer wave packets in configuration space through this
discrete Fourier representation.
The truncation parameters are chosen as $
\mathcal{L}=\mathcal{S}=10$, providing $2\mathcal L +1=21$ basis states for each pointer degree of freedom.
The qualitative measurement dynamics is already reproduced for much
smaller truncations (e.g., $\mathcal{L}=\mathcal{S}=1$); the chosen value
is adopted to obtain better spatial resolution of the pointer packets in configuration~space.

For an initial Gaussian packet of spatial width $\sigma_0$, the~time evolution leads to
\linebreak  $\sigma(t)=\sigma_0\sqrt{1+\left(\hbar t/2 m_e \sigma_0^2\right)^2}$.
Pointer dynamical stability therefore requires the dispersion time scale
$
\tau_d \sim \frac{2 m_e \sigma_0^2}{\hbar}$
to be much larger than the total interaction and simulation time, $\tau_d \gg T_{\text{sim}}$, and, simultaneously, that the spatial extension of the apparatus be sufficiently large so that the wave packets remain well inside the physical domain during the evolution, i.e.,~$
L_{y,z} \gg \sigma_{y,z}(t).$
These conditions can be summarized by the hierarchy
\begin{equation}
L_{y,z} \gg \sigma_{y,z}(t) \gg \sqrt{\frac{\hbar T_{\text{sim}}}{2 m_e}}.
\end{equation}

In the present implementation we choose $m_y=m_z=m_e$. For~a total simulation time of approximately one Rabi period, $T_{\text{sim}} \approx  T_R$, the~condition $\tau_d \gg T_{\text{sim}}$ implies
\begin{equation}
\sigma^{y,z}_0 \gg \sqrt{\frac{\hbar T_{\text{sim}}}{2 m_e}} \sim 10^{-8}\,\mathrm{m}.
\end{equation}

The width $\sigma^{y,z}_0$ of the Gaussian wave packet in real space is approximately related to the width of the Gaussian coefficients in the conjugate mode space, $\sigma_{k_y,k_z}$, through
$
\sigma^{y,z}_0 \approx L_{y,z}/2\pi\sigma_{k_y,k_z}$.
The parameters $\sigma_{k_y,k_z}$ are chosen such that truncation effects from the finite basis are negligible. In~particular, we use
$
\sigma_y=\sqrt{\mathcal{L}}, \sigma_z=\sqrt{\mathcal{S}},
$
which ensures that the Gaussian distribution of coefficients is well contained within the truncated interval of modes and produces a smooth Gaussian wave packet in real~space.

With the choice $L_{y,z}=10^{-6}\,\mathrm{m}$, this gives an initial spatial width $
\sigma^{y,z}_0 \sim 10^{-7}\,\mathrm{m}.
$
The resulting parameters satisfy the hierarchy $
L_{y,z} \gg \sigma^{y,z}(t) \gg \sqrt{\hbar T_{\text{sim}}/2 m_e},$
ensuring negligible spreading during the simulated time window and allowing the separated pointer branches to remain dynamically~distinguishable.

%\section{Computation of the Bohmian~Trajectories}
\section[\appendixname~\thesection. Computation of the Bohmian~Trajectories]{Computation of the Bohmian~Trajectories}
\label{ap:trajectories}

We compute here the Bohmian velocities of the measured light--matter system. The~Bohmian velocities of the non-measured simulations can be obtained straightforwardly. We consider the full system with five degrees of freedom $(x_1,x_2,q,y,z)$ describing the two electrons, the~cavity mode and the two pointer variables of the measuring devices. The~Bohmian velocities of the non-measured simulations can be obtained~straightforwardly.  

The wave function $\Phi(x_1,x_2,q,y,z,t)$ satisfies the Schrödinger in \eqref{schomeasure} in the manuscript, written here as $i\hbar\partial_t\Phi-H\Phi=0$. To~obtain the associated conserved current, we look for a continuity equation. Following the standard procedure we subtract the complex conjugate equation multiplied by $\Phi$ from the original equation multiplied by $\Phi^*$,
\begin{equation}
\Phi^*(i\hbar\partial_t\Phi-H\Phi)
-
(i\hbar\partial_t\Phi-H\Phi)^*\Phi=0 .
\end{equation}
The terms involving a time derivative can be compactly written as $i\hbar\frac{\partial |\Phi|^2}{\partial t}$.  All potential terms, say $f$, cancel because they appear without spatial derivatives, \mbox{$\Phi^* f \Phi-\Phi f \Phi^*=0$}. The~remaining terms with spatial  derivative can be rearranged as total derivatives. For~instance,
\begin{equation}
\Phi^*\partial_x^2\Phi-\partial_x^2\Phi^*\Phi
=
\partial_x
\left(
\Phi^*\partial_x\Phi-\Phi\partial_x\Phi^*
\right).
\end{equation}
The mixed derivatives appearing in the measurement interaction can also be written as total derivatives,
\begin{equation}
\Phi^*\partial_y\partial_x^2\Phi
-
\partial_y\partial_x^2\Phi^*\Phi
=
\partial_x
\left(
\Phi^*\partial_{yx}^2\Phi+\Phi\partial_{yx}^2\Phi^*
\right)
-
\partial_y
\left(
\partial_x\Phi^*\partial_x\Phi
\right).
\end{equation}
After rearranging all terms we obtain the following continuity equation:
\begin{equation}
\frac{\partial |\Phi|^2}{\partial t}
+
\frac{\partial J_{x_1}}{\partial x_1}
+
\frac{\partial J_{x_2}}{\partial x_2}
+
\frac{\partial J_q}{\partial q}
+
\frac{\partial J_y}{\partial y}
+
\frac{\partial J_z}{\partial z}
=0 .
\end{equation}
The probability currents define the Bohmian velocity fields through the expression $J_i = |\Phi|^2 v_i$. So finally, we get
\begin{eqnarray}
v_{x_1}&=&
\frac{
-i\frac{\hbar}{2m_e}
(\Phi^*\partial_{x_1}\Phi-\Phi\partial_{x_1}\Phi^*)
-
\mu(t)\frac{\hbar^2}{2m_e}
\left(
\Phi^*\partial_{yx_1}^2\Phi+\Phi\partial_{yx_1}^2\Phi^*
\right)
}{|\Phi|^2},
\\
v_{x_2}&=&
\frac{
-i\frac{\hbar}{2m_e}
(\Phi^*\partial_{x_2}\Phi-\Phi\partial_{x_2}\Phi^*)
-
\mu(t)\frac{\hbar^2}{2m_e}
\left(
\Phi^*\partial_{zx_2}^2\Phi+\Phi\partial_{zx_2}^2\Phi^*
\right)
}{|\Phi|^2},
\\
v_q&=&
\frac{
-i\frac{\omega}{2}
(\Phi^*\partial_q\Phi-\Phi\partial_q\Phi^*)
}{|\Phi|^2},
\\
v_y&=&
\frac{
-i\frac{\hbar}{2m_y}
(\Phi^*\partial_y\Phi-\Phi\partial_y\Phi^*)
+
\mu(t)\frac{\hbar^2}{2m_e}
\left(
\partial_{x_1}\Phi^*\partial_{x_1}\Phi
\right)
}{|\Phi|^2},
\\
v_z&=&
\frac{
-i\frac{\hbar}{2m_z}
(\Phi^*\partial_z\Phi-\Phi\partial_z\Phi^*)
+
\mu(t)\frac{\hbar^2}{2m_e}
\left(
\partial_{x_2}\Phi^*\partial_{x_2}\Phi
\right)
}{|\Phi|^2}.
\end{eqnarray}

% ACS format

\end{document}